\documentclass[aps,pre,11pt,amsfonts,amsmath,amssymb]{revtex4-1}

\usepackage{amssymb}
\usepackage{graphicx}
\usepackage{dcolumn}
\usepackage{bm}
\usepackage{appendix}

\usepackage{algorithm}
\usepackage[noend]{algpseudocode}

\newcommand{\Cov}[2]{\mathbb{C}\mathrm{ov}\left(#1,#2\right)}
\newcommand{\EE}[1]{\mathbb{E}\left[#1\right]}
\newcommand{\OO}[1]{O\left(#1\right)}
\newcommand{\PP}[1]{\mathbb{P}\left[#1\right]}

\newcommand{\cA}{{\mathcal A}}

\begin{document}

\preprint{APS/123-QED}

\title{Continuous cascades in the wavelet space as models for synthetic turbulence}

\author{Jean-Fran\c{c}ois Muzy}
\email{muzy@univ-corse.fr}
\affiliation{SPE UMR 6134, CNRS, Universit\'e de Corse, 20250 Corte, France}

\date{\today}

\begin{abstract}
We introduce a wide family of stochastic processes that are obtained as
sums of self-similar localized ``waveforms'' with multiplicative intensity 
in the spirit of the Richardson cascade picture of turbulence. We establish the convergence 
and the minimum regularity of our construction. We show that
its continuous wavelet transform is characterized by stochastic self-similarity and multifractal scaling properties. 
This model constitutes a stationary, "grid free", extension of $\cal W$-cascades introduced in the past 
by Arneodo, Bacry and Muzy using wavelet orthogonal basis.
Moreover our approach generically provides multifractal 
random functions that are not invariant by time reversal and therefore is able to account for skewed multifractal models and for the so-called ``leverage effect".
In that respect, it can be well suited to providing 
synthetic turbulence models or to reproducing the main observed 
features of asset price fluctuations in financial markets.

\end{abstract}


\maketitle
\section{Introduction}
The goal of modeling the observed ``random'' fluctuations of the velocity field and the intermittent character of the small scale dissipated energy in fully developed turbulent flows has played a critical role in the development of mathematical concepts around multifractal processes. In particular, 
random multiplicative cascades first considered by the Russian
school \cite{kol62,ob62,NovSt64} and subsequently developed by B. Mandelbrot \cite{Man74a,Man74b} in order to mimic 
the energy transfer from large to small scales \cite{Fri95} 
represent the paradigm of multifractal random distributions.
They are the basis of a lot of mathematical work
and have led to a wide number of applications
far beyond the field of turbulence.

Mandelbrot cascades ($\cal M$-cascades) mainly consist 
in building a random measure by using a multiplicative iterative rule. One starts with
some large interval with a constant measure density $W_0$ and splits
this interval in two equal parts. The measure density in the
left and right half parts are obtained by multiplying $W_0$ by respectively two independent, identically distributed 
positive factors (say $W_1^L$ and $W_1^R$). This operation is then repeated 
independently on the two sub-intervals and so on, {\em ad infinitum}, in
order to converge towards a singular measure which properties
have been studied extensively (see e.g. \cite{Man74b,KaPe76,Gui87,Bar04}).
The main disadvantage of $\cal M$-cascades is that they involve a
specific scale ratio ($s=2$ in general) and are limited to live in the starting interval.
This lack of stationarity and continuous scale invariance
is obviously not suited to accounting for natural phenomena. 
In order to circumvent these problems, continuous extensions of Mandelbrot cascades were proposed by first Barral and Mandelbrot \cite{BaMan02} and later by Bacry and Muzy \cite{MuBa02,BaMu03}. The idea under these
constructions is to replace the discrete multiplicative density
$\prod_{i=1}^n W_i = e^{\sum_{i=1}^n \ln W_i}$ by $e^{\omega_\ell(t)}$ where $\omega_{\ell}(t)$ is an infinitely divisible noise chosen with a logarithmic correlation function designed to mimic the tree-like (in general a dyadic tree) structure underlying $\cal M$-cascades \cite{ArnBaMaMu98,MuDeBa00,sm01}.

The above definitions of random cascade measures led to a large number of extensions notably in order to construct stochastic processes with some predefined regularity properties.
The most popular approach was initiated by Mandelbrot (see e.g. \cite{ManFisCal97}) and consists in compounding a self-similar stochastic process like the (fractional-) Brownian motion $B_H(t)$ with a ``multifractal time" 
$M(t)$, a multifractal measure of the interval $[0,t]$ as provided by a continuous cascade. 
Such a compounded process $B_H(M(t))$ inherits the multifractal
scaling properties of $M(t)$ with a "main" regularity that corresponds to the self-similar process $B_H(t)$. 
An alternative but related approach, was initially proposed 
with the construction of ``the Multifractal Random Walk" (MRW) in \cite{MuDeBa00,BaDeMu01} and that consists of interpreting $M(dt)$ as the local variance of a fractional Brownian motion $B_H(t)$. As 
emphasized in \cite{MuBa02,lud08,accp09}, this amounts to constructing a multifractal process as (the limit of)  a stochastic integral like $\int e^{\omega_{\ell}(t)} dB_H(t)$.
This point of view is inspired from classical stochastic volatility models of financial markets that aim at accounting for the observed bursts in price fluctuations by the random nature of the variance of an underlying Gaussian law. It can also be understood as a formalization of the so-called "Kolmogorov refined similarity hypothesis" \cite{Fri95} according to which the velocity fluctuation $\delta v$ and 
the local rate of dissipated energy $\varepsilon$ are related as,
in its Lagrangian version \cite{rshLag09}, $\delta_\tau v \sim \varepsilon_\tau^{1/2} dW_\tau$, $dW_\tau$ being a noise such that $\EE{dW_\tau^2} = \tau$. Accordingly, the multifractal properties 
of the velocity field directly come from those of the local
dissipation field $\varepsilon$, a multifractal cascade interpreted as a ``local variance''. 
The main advantage of continuous cascades and their associated random processes is that they provide a large class of parsimonious models with stationary increments and continuous stochastic scale invariance properties. From a practical point of view, they are easy to calibrate from data and can provide simple analytical or numerical solutions to many statistical problems. However, the aforementioned methods to derive a multifractal process from a multifractal measure lack of flexibility and in particular hardly allow one to describe
processes that are not invariant by time reversal. This means that 
the increment distribution skewness (i.e. non vanishing third order moments of the increments at different scales) as observed in turbulence or the leverage effect (i.e. some causal, asymmetric relationship between increment signs and increment amplitudes) as observed in financial time series cannot be captured easily within this framework. All former attempts to address this issue were based on studying a random noise like $e^{\omega_\ell(t)} \Delta B_H(t)$ for a specific cross-covariance between $\omega_\ell$ and $\Delta B_H$ \cite{PocBou02,BaDuMu12}. However, as discussed in \cite{BaDuMu12}, this approach leads to a non degenerated limiting process only in some restricted range $H > 1/2$ and therefore cannot be used as a model for financial markets ($H \simeq 1/2$) or turbulence ($H \simeq 1/3$).
More recently, Chevillard et al. \cite{CheGaRhoVa17} introduced a new model which obeys to multifractal scaling with a (signed) third order structure function that behaves as a linear function, like in turbulence.
This model is defined within the context of Gaussian Multiplicative Chaos (i.e. a Gaussian version of $\omega_{\ell}(t)$) and mainly consists in considering a fractional integration of a product like $\omega' e^{\omega} dB$ where $\omega'$ represents a peculiar intermittent version of $\omega$, independent of $\omega$ but correlated with the white noise $dB$. The authors have shown that
their construction leads to a properly defined intermittent skewed random process. However, we can point out that the model of Chevillard et al. is far from being simple, involves cumbersome computations and is not characterized by self-similar properties, it follows a multifractal scaling only in the limit of small scales.

In this paper, we propose a different path to solve 
the challenging issue of mixing multiscaling properties and non-invariance by time reversal.
Beyond this question, our construction also offers an appealing alternative way to build a large class of multifractal functions
with well controlled scaling (or regularity) properties and that are
characterized by new features as compared to former constructions.
Our approach is directly inspired from the random cascade
picture originally proposed by Richardson \cite{Rich22} under which, in turbulent flows, large eddies are stretched and broken into smaller ones to which they transfer a fraction of their energy and so on, up to the dissipation scale. Instead of trying to capture the velocity field intermittency from the final state of such a cascade, i.e., the dissipation field (which is a multifractal measure), the previous scenario suggests that one could describe the overall flow as a {\em superposition} of coexisting structures at all scales correlated with each other by a cascading intensity (energy).
This viewpoint of decomposing a function as a weighted sum of waveforms (the "eddies") at different scales precisely corresponds to a wavelet transform representation \cite{Mallatbook}. The construction of discrete multiplicative cascades along the wavelet tree associated with orthogonal wavelet basis has been already 
proposed two decades ago by Arneodo, Bacry and Muzy \cite{ArnBacMuz98}.
These "$\cal W$-cascades" have proven to be an appealing alternative to $\cal M$-cascade based constructions in order to directly build multifractal stochastic processes \cite{ArnBaMaMu98,BarSeu05}. They also provided a suitable approach to extent the framework of stochastic self-similarity and build new random functions with a non trivial spectrum of oscillating singularities \cite{ABJM97,ABJM98}.
Our goal in this paper is thus to construct an extension of 
$\cal W$-cascades in order to get rid of the restraining grid structure of wavelet orthogonal basis that prevent, very much like grid-bounded $\cal M$-cascades, $\cal W$-cascades from being able to simply account for stationarity and continuous scale invariance. For that purpose, we just consider a continuous sum over space $t$ and scales $\ell$ of ``sythetizing wavelets'' weighted by a factor that is precisely given by the stochastic density involved in continuous cascades, $e^{\omega_{\ell}(t)}$. We show that such a construction allows us to obtain well defined processes with a large flexibility on their scaling and regularity properties. Moreover it generically leads to skewed processes and is able to reproduce the leverage effect for specific shapes of synthetizing wavelets. In that respect, they can naturally be invoked as models for synthetic turbulence or calibrated to account for asset fluctuations in financial markets. 
Various numerical examples are provided throughout the paper in order to illustrate our 
purpose and notably to show the model ability to reproduce many of the 
observed features of the longitudinal 
velocity field  fully developed turbulence experiments.

The paper is structured as follows: in section \ref{sec:wcasc} we recall the main lines of continuous 
cascades (Sec. \ref{sec:c-casc}) and $\cal W$-cascades (Sec. \ref{sec:w-casc}) constructions. 
This notably allows us to introduce the process $\omega_{\ell}(t)$ and review its main statistical properties. Our new constrution of continuous cascades in the wavelet plane 
($\cal C W$-cascades) are introduced in section \ref{sec:wc-cascades}.
After the definition and the statement of a weak convergence result, 
we provide some numerical examples (Sec. \ref{sec:wc-cascade_def}). In Sec. \ref{sec:wc-cascades_wavelets}, we study their wavelet transform and establish the almost sure minimum regularity of their paths. Scaling and self-similarity
properties of this new class of processes are studied
in section \ref{sec:scaling}. We prove that their structure functions are characterized by a power-law behavior with some non-linear $\zeta_q$ spectrum of scaling exponents (Sec. \ref{sec:wc-cascades_ss}) and
study the correlation functions of the absolute increments (Sec. \ref{sec:wc-cascades_corr}). We finally
study the properties of $\cal CW$-cascades as respect to time reversal, notably the behavior of the skewness and the leverage function (Sec. \ref{sec:cw-cascades_skew}). 
Concluding remarks and prospects for future research are given in section \ref{sec:conclusion} 
while technical material and proofs are provided in appendices. 

\section{Continuous cascades and $\cal W$-cascades}
\label{sec:wcasc}
This section contains a brief review of the notions
of ``continuous'' and ``wavelet'' multiplicative
cascades. The first ones were introduced as stationary, self-similar singular measures with log-infinitely divisible multifractal properties while
$\cal W$-cascades are the wavelet transform counterparts
of Mandelbrot discrete multiplicative cascades. As 
explained in Sec. \ref{sec:wc-cascades},
both constructions will be mixed in order to build
continuous wavelet cascade models.

\subsection{Log-infinitely divisible continuous cascades}
\label{sec:c-casc}
Continuous random cascades are stochastic measures $M(t)$ introduced few years ago (\cite{BaDeMu01,SchMar01,BaMan02,MuBa02,BaMu03}) in order to extend, within a stationary and grid-free framework,
the Mandelbrot discrete multiplicative cascades. 
Such a measure $M(t) = \int_0^t dM_u$ 
can have exact multifractal scaling properties in the sense that it satisfies, for a given $T >0$, $\forall t \leq T$:
\begin{equation}
\label{eq:scaling_cascades}
 \EE{M(t)^q} = K_q \left(\frac{t}{T} \right)^{\zeta_q}
\end{equation}
where $\zeta_q$ is a non-linear
concave function called the multifractal spectrum (or the spectrum of structure functions scaling exponents in the context of turbulence \cite{Fri95,Ar_Turb96}) and $K_q$ is a (eventually infinite) prefactor
corresponding the the $q$-order moment at scale $t=T$.

In \cite{MuBa02, BaMu03} (see also \cite{BaMan02}) $dM(t)$ is obtained as the (weak) limit, when $\ell \to 0$ of the measure with density:
\begin{equation}
\label{eq:def_M}
  dM_\ell(t)  = e^{\omega_{\ell}(t)} dt
\end{equation}
where $e^{\omega_{\ell}(t)}$ is a stationary log-infinitely divisible process
representing the continuous cascade from scale $T$ to scale $\ell$. Its precise definition
and main properties, reviewed in Appendix \ref{app:cont_casc}, notably imply that it satisfies
(thanks to Eqs. \eqref{rhoexact} and \eqref{charf}):
\begin{equation}
\label{eq:def_phi}
\EE{e^{q \omega_{\ell}(t)}} = \left(\frac{T}{\ell}\right)^{\phi(q)}
\end{equation}
where $\phi(q)$ is the cumulant generating function associated with 
an infinitely divisible law as provided by the celebrated Levy-Khintchine Theorem \cite{Fel71}. Moreover, $\omega_{\ell}(t)$ verifies, $\forall s < 1, u \in [0,T]$,
\begin{equation}
\label{eq:ss-omega}
\omega_{s \ell}(s u)  \stackrel{\cal L}{=}  \Omega_s + \omega_\ell(u)
\end{equation}
where $\stackrel{\cal L}{=}$ means that the two processes have the same 
finite dimensional distributions of any order and $\Omega_s$ is a random variable independent of the process $\omega_\ell(t)$ with the same distribution as $\omega_{s T}(t)$.
The multifractal scaling \eqref{eq:scaling_cascades} follows since
Eq. \eqref{eq:ss-omega} entails
\begin{equation}
  M(st_0) = \lim_{\ell \to 0} \int_0^{s t_0} e^{\omega_{s \ell}(u)} du \stackrel{\cal L}{=}  se^{\Omega_s} \lim_{\ell \to 0}  \int_0^{t_0} e^{\omega_{\ell}(u)} du  =  se^{\Omega_s} M(t_0)
\end{equation}
and thus by choosing $t_0=T$ and $s = t/T$, we have
$$
 \EE{M(t)^q} = s^q \EE{e^{q\Omega_{s}}} \EE{M(T)^q}
$$
that leads, using Eq. \eqref{eq:def_phi}, to Eq. \eqref{eq:scaling_cascades} with
$$
 \zeta_q = q-\phi(q) \; .
$$
As discussed in the introduction, a large class of multifractal processes $X(t)$ with stationary increments can be obtained from a multifractal measure $M(t)$. One can, as suggested by Mandelbrot \cite{ManFisCal97,Man99}, compound a self-similar process $B_H(t)$ (e.g., a fractional Brownian motion) by the non-decreasing function $M(t)$ so that:
$$
  X(t) = B_H(M(t)) \; .
$$
Within this approach $M(t)$ is referred to as the "multifractal time".
Since $B_H(st) \stackrel{\cal L}{=} s^H B_H(t)$, one has
$$
  X(st) \stackrel{\cal L}{=} s^H e^{H \Omega_s} X(t)
$$
that entails, given the increment stationarity of $X(t)$, the multiscaling of the structure functions:
$$
 \EE{|X(t+\tau)-X(t)|^q} = C_q \tau^{\zeta_q}
$$
with $\zeta_q = qH-\phi(qH)$. Another approach initiated in \cite{MuDeBa00,BaDeMu01} (see
also \cite{MuBa02}) is to consider the limit $\ell \to 0$
of the sequence $X_\ell(t) =\sum_{k=1}^{t/\ell} \ell^{\alpha} e^{\omega_\ell(k)} dW_k$ where $dW_k$ is a fractional
Gaussian noise and $\alpha$ an appropriate constant chosen to 
ensure the convergence (in some specific sense) of the series. 
This amounts to interpreting $e^{\omega_\ell(k)}$ as a stochastic
variance or, in the field of mathematical finance, a "stochastic volatility". Related
constructions consist in considering  stochastic integrals like $\int e^{\omega_{\ell}(t)} dB_H(t)$, 
where $B_H(t)$ is a fractional Brownian motion \cite{MuBa02,lud08,accp09}.
All these approaches were extensively used and studied 
in the literature as paradigms of multifractal 
processes satisfying exact stochastic scale-invariance properties and also considered
as toy models for applications like turbulence or financial time series.
However, as recalled in the introduction, they mainly involve separately the construction
of a multifractal measure and a self-similar process and do not consist in directly
building the random process $X(t)$ with some specific properties. It then results
a lack of flexibility in the obtained features, in particular,
as emphasized previously and discussed in \cite{PocBou02,BaDuMu12,CheGaRhoVa17}, skewed statistics and "leverage effect" cannot be obtained in a fully satisfactory way through these approaches. As reviewed below, wavelet cascades offer an interesting alternative in the 
sense that they do not rely on any preset self-similar process and consists in directly
building $X(t)$ with a control of its scaling or regularity properties.

\subsection{$\cal W$-cascades}
\label{sec:w-casc}
In Ref. \cite{ArnBacMuz98}, the authors introduced the so-called 
$\cal W$-cascades as the natural 
transposition of Mandelbrot $\cal M$-cascades within the framework
a orthogonal wavelet transform. This allows one to construct multifractal processes or distributions with a precise control of their regularity properties.
The idea is to build a new class of (multi-)fractal functions $Z(t)$ from their explicit representation over 
a wavelet basis:
\begin{equation}
\label{eq:wcascade1}
Z(t) = \sum_{j=0}^\infty \sum_{k=0}^{2^{j-1}} c_{j,k} \psi_{j,k}(t)
\end{equation}
where $\psi_{j,k}(t) = 2^{j} \psi(2^j t-k)$ and $\{ \psi_{j,k}(t) \}_{j,k \in \mathbb Z}$ constitutes an orthogonal 
wavelet basis of the interval. The wavelet coefficients $c_{j,k} = \int Z(u) \psi_{j,k}(u) du $ are chosen according to the multiplicative cascade rule \cite{ArnBacMuz98}:
\begin{equation}
\label{eq:wcascade2}
c_{j,2k} = W_0 \; c_{j-1,k}, \; \;  c_{j,2k+1} = W_1 \; c_{j-1,k}
\end{equation}
with $W_0$ and $W_1$ are i.i.d. copies of a real valued random variable $W$.
We see that, if $W$ is a positive random variable, the law of the wavelet coefficient $c_{j,k}$ is precisely
given by the law of the density of a $\cal M$-cascade at construction step $j$:
\begin{equation}
\label{eq:wcascade3}
 c_{j,k} = \prod_{m=0}^{j} W_m
\end{equation}
It has been shown in \cite{ArnBacMuz98} that, under some mild condition on the law of $W$, $Z(t)$ is a well
defined multifractal process with almost surely Lipschitz regular paths.
Such a construction of multifractal functions associated with specific 
random wavelet series has been extended recently 
by Barral and Seuret \cite{BarSeu05}. If $M$ stands for a Mandelbrot 
multifractal measure (a $\cal M$-cascade)
constructed from an iterative random multiplicative rule as described in the introduction, these authors considered a wavelet random series like \eqref{eq:wcascade1} but where $c_{j,k}$ is given by the measure of the associated dyadic interval $I_{j,k}=[k2^{-j},(k+1)2^{-j}]$:
$$
c_{j,k} = 2^{-j \alpha} M(I_{j,k}) \; .
$$
Barral and Seuret proved that the scaling and regularity
properties of $Z(t)$ are directly inherited from those
of $M(t)$. Moreover, they have shown that, under specific conditions, replacing $\prod_{i=1}^j W_i$ of $\cal W$-cascades by the limit measure of the associated dyadic interval, $M(I_{j,k})$ does not change the multifractal
properties (see the Sec. \ref{sec:wc-cascades_ss} for more details). 

Unlike the constructions of multifractal processes 
based on multifractal measures, $\cal W$-cascades allow one to directly build multifractal processes without the need for any additional self-similar process.
However, very much like $\cal M$-cascades, they do involve a dyadic tree and a finite time interval 
that can hardly be used to fit most of experimental situations.
For that reason, in the same manner as $\cal M$-cascades have been extended to log-infinitely divisible 
continuous cascades, we aim at defining a continuous version of $\cal W$-cascades.

\section{Continuous $\cal W$-cascades}
\label{sec:wc-cascades}
In this section, we introduce the new class of models we consider in the paper. 
Our goal is to extend the previously described $\cal W$-cascades
to a grid-free background. The main idea is to replace the discrete sum \eqref{eq:wcascade1}
by a continuous sum over space and scales and the discrete product in Eq. \eqref{eq:wcascade3} at 
scale $\ell = 2^{-j}$ by its ``continuous" (i.e. stationary and "grid free") counterpart $e^{\omega_\ell(t)}$  
described in Sec. \ref{sec:c-casc} and Appendix \ref{app:cont_casc}. 

\subsection{Definition and numerical illustrations}
\label{sec:wc-cascade_def}
\subsubsection{Definition and convergence}
Let us define a stochastic process $X(t)$ as a (continuous) 
sum of localized waveforms (i.e. "wavelets") of size $\ell$ and which 
intensity is given by the class of stationary, log-infinitely divisible 
process $e^{\omega_{\ell}(t)}$ used in the definition of continuous cascades of Section \ref{sec:c-casc}.

Let $H>0$, $\ell >0$ and consider the following (well defined) integral:
\begin{equation}
\label{eq:def_X}
  X_\ell(t) = \int_{\ell}^T \! s^{H-2} ds  \! \int_{-\infty}^{+\infty}   e^{\omega_{s}(b)} \varphi \left(\frac{t-b}{s} \right)  db 
\end{equation}
where $\omega_{s}(u)$ is the infinitely divisible process defined 
in Section \ref{sec:c-casc} and $\varphi(t)$ is a wavelet 
that can be chosen as a square integrable smooth function with compact support (e.g, the interval $[-{1 \over 2},{1 \over 2}]$) 
and which is sufficiently oscillating so that its 
first $N$ moments vanish. Hereafter, we will referred to this wavelet as the ``synthetizing wavelet". 
As emphasized below, this amounts, in some sense, to interpreting $s^H e^{\omega_{\ell}(t)}$
as the continuous wavelet transform of $X(t)$ at time $t$ 
and scale $\ell$, Eq. \eqref{eq:def_X} corresponding to the continous 
wavelet reconstruction formula.
Let us remark that, if $\phi(1) < \infty$,  one can, without loss of generality (since it simply consists in redefining the parameter $H$), always assume 
that in Eq. \eqref{eq:def_X}, $\omega_s(t)$ is chosen such that
\begin{equation}
\label{eq:phi1}
 \EE{e^{\omega_s(t)}} = 1 \Leftrightarrow \phi(1) = 0 \; .
\end{equation}

In Appendix \ref{App:conv_C0}, we establish the weak convergence 
of $X_{\ell}(t)$ in the space of continuous functions when $\ell \to 0$. Namely, we show that the weak limit
\begin{equation}
\label{eq:lim_X}
 \lim_{\ell \to 0} X_\ell(t) = X(t)
\end{equation}
exists as a continuous function provided:
\begin{equation}
\label{eq:conv-condition}
  H > \frac{\phi(2)}{2} \; .
\end{equation}
We will call such a limit $X(t)$ a ``continuous wavelet cascade'' process or a  ${\cal CW}$-cascade.
Let us notice that the condition \eqref{eq:conv-condition} is precisely
the analog of the condition for $L^2$ convergence of $\cal W$-cascades established by Arneodo et al. (condition $H1$ of Proposition 1 in \cite{ArnBacMuz98}).

\begin{center}
	\begin{figure}[h]
		\includegraphics[width=0.9\textwidth]{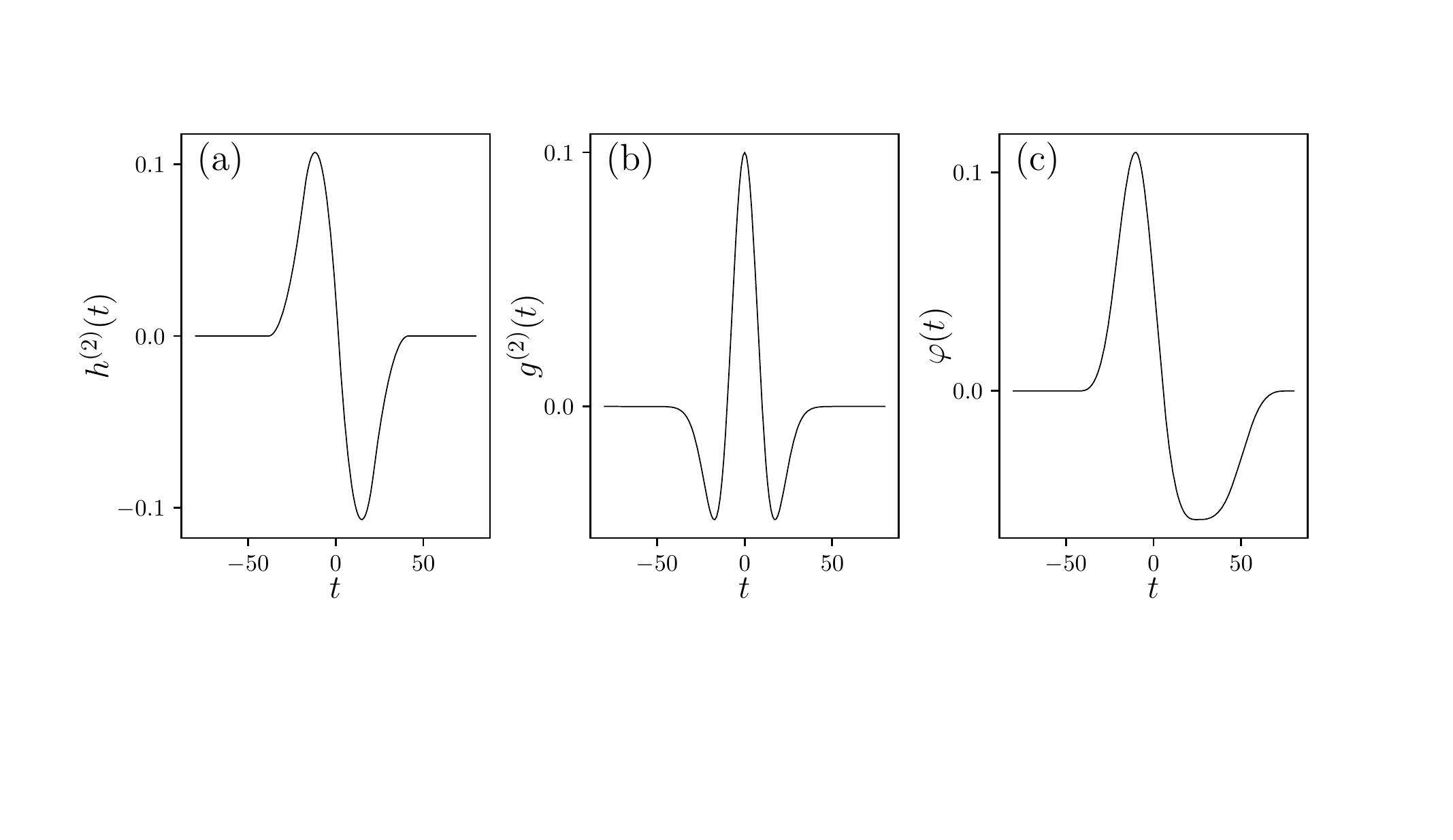} 
		\vspace{-3cm}
		\caption{Examples of synthetizing wavelets $\varphi(t)$. (a) The compactly supported 
			odd	function $h^{(2)}(t)$ (Eq. \eqref{eq:smH}). (b) The "Mexican hat", i.e. the even wavelet $g^{(2)}(t)$ (Eq. \eqref{eq:gauss_deriv}). (c) A non symmetric wavelet obtained by smoothing an asymmetric version 
			of the Haar wavelet.}
		\label{fig_wavelets}
	\end{figure}
\end{center}

\begin{center}
	\begin{figure}[h]
		\includegraphics[width=0.7\textwidth]{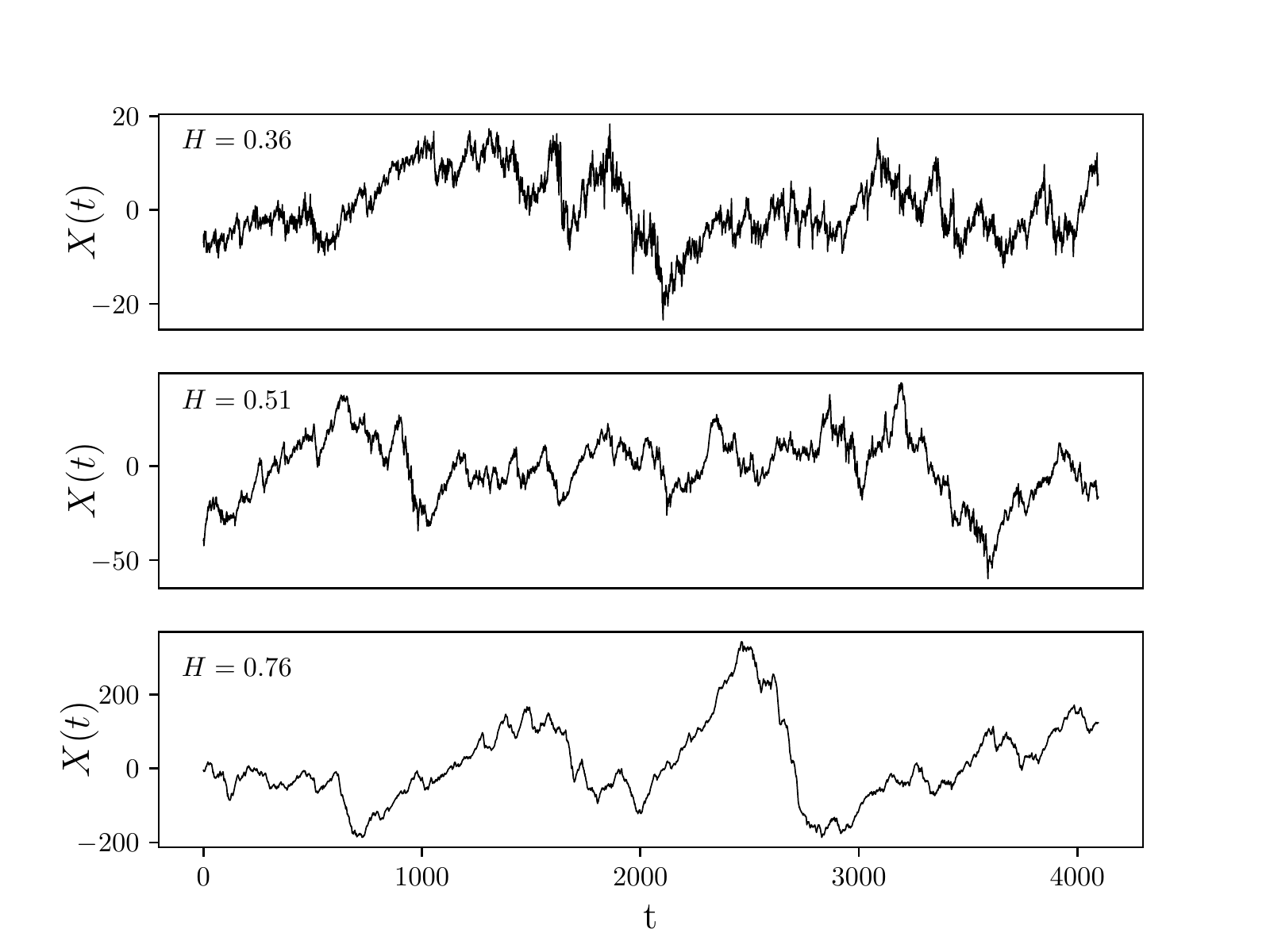}
		\caption{Sample paths of continuous log-normal $\cal W$-cascades $X(t)$ corresponding to  
			3 different values of the parameter
			$H$. In all cases, the intermittency coefficient  and the integral scale were chosen 
			to be respectively $\lambda^2 = 0.025$ and $T = 1024$ and the synthetizing wavelet the $C^2$ version
			of the Haar wavelet, $h^{(3)}(t)$.}
		\label{fig_paths}
	\end{figure}
\end{center}

\subsubsection{Numerical examples} 
In numerical experiments, one has to choose the infinitely divisible law of the process $\omega_{\ell}(t)$,
the regularity parameter $H$, the integral scale $T$ and the synthetizing wavelet $\varphi(t)$.
The simulation procedure simply consists in a discretization of 
Eq. \eqref{eq:def_X}. Its main lines are described in Appendix \ref{app:simus}. 

All the examples provided in the paper are involving log-normal cascades which are the simplest 
ones to handle and that involve a single variance parameter $\lambda^2$. This means that $\omega_{\ell}(t)$
is a Gaussian process with a covariance function given by expression \eqref{rhoexact}.
In that respect, thanks to condition \eqref{eq:phi1}, the function $\phi(q)$ simply reads:
\begin{equation}
\label{eq:phiq_ln}
  \phi(q) = \frac{\lambda^2}{2}q(q-1) \; .
\end{equation}
and the condition \eqref{eq:conv-condition} simply becomes
$\lambda^2 < 2 H$. 
Let us remark that this condition appears to be 
less restrictive than the condition of existence of a MRW $\lambda^2 < 1/2$ \cite{BaDeMu01,DelourPHD} and it is a priori possible to build ${\cal CW}$-cascades with a large intermittency coefficient.

Among the choices we made for the synthetizing wavelets $\varphi$  
there is the class of smooth variants of the Haar wavelet:
\begin{equation} 
\label{eq:smH}
   h^{(n)}(t) = \left(\mathbb{I} \underset{n}{\star \ldots \star} \mathbb{I} \star h \right)(t)
\end{equation}
where $\mathbb{I}(t)$ is the indicator function of the interval $[0,1/2)$, $h(t) = \mathbb{I}(t+1/2)- \mathbb{I}(t)$ is the Haar wavelet and $\star$ stands for the convolution product. Notice that $h^{(n)}(t)$
has one vanishing moment and is of class $C^{n-1}(\mathbb{R})$.
We also use wavelets in the class of the derivatives of the Gaussian function:
\begin{equation}
\label{eq:gauss_deriv}
  g^{(n)}(t) = \frac{d^n}{dt^n} e^{-\frac{x^2}{2}} \; .
\end{equation}
Some examples among these wavelet classes are plotted in Fig. \ref{fig_wavelets} (dilation and normalization factors are arbitrary).
Note that one can also consider asymmetric wavelets
as the one illustrated in the right panel of Fig. \ref{fig_wavelets} that is contructed by asymmetrizing and smoothing the Haar wavelet
(see Sec. \ref{sec:cw-cascades_skew} for an usage of asymmetric wavelets).

Some examples of sample paths of $X(t)$ are plotted in Figs. \ref{fig_paths}.
All processes were generated using a Gaussian process $\omega_{\ell}(t)$
with the intermittency coefficient $\lambda^2 = 0.025$ and the integral scale $T = 2048$.
The synthetizing function was chosen to be $\varphi(t) = h^{(3)}(t)$. 
The scaling parameter $H$ has been chosen to be respectively $H=0.36$, $H=0.51$ and $H = 0.76$ from top to bottom.
One can see that $H$ directly controls 
the global regularity of the paths, the larger $H$, the more regular the path of $X(t)$ is. This is the analog of the parameter $H$ of the fractional Brownian motion \cite{Taqqu}. It is noteworthy that despite $X(t)$ appears to be a process with
zero mean stationary increments, unlike all constructions proposed so far for mulifractal stationary processes, it does not involve any supplementary "white" or "colored" noise,
like the Gaussian white noise, since it is only based on the "intensity" process $e^{\omega_{\ell}(t)}$ as a source of randomness (see the remark at the end of the next section).

\subsection{Wavelet transform, global regularity and reconstruction formula} 
\label{sec:wc-cascades_wavelets}
In order to study and characterize the limiting process $X(t)$, 
one can compute its wavelet transform \cite{Mallatbook}. 
If $\psi(t)$ stands for some analyzing wavelet,
let us introduce the kernel $K_{\varphi,\psi}(x,s)$ defined as:
\begin{equation}
K_{\varphi,\psi}(x,s) = s^{-1} \int \varphi \left(t \right) \psi \left(\frac{t-x}{s} \right) dt \; .
\end{equation}

The wavelet transform of $X(t)$ can then be simply expressed as:
\begin{equation}
\label{eq:wtX}
W(x,a) = a^{-1} \int X(t) \psi \left( \frac{t-x}{a} \right) dt = \int_0^T s^{H-2} ds \! \! \int_{-\infty}^{+\infty} db \;  e^{\omega_{s}(b)} K_{\varphi,\psi}\left(\frac{x-b}{s},\frac{a}{s} \right)   
\end{equation}

It is easy to show that if $\psi$ has more than $N$ vanishing moments one 
has $K_{\varphi,\psi}(x,s) \sim s^{N}$ when $s \ll 1$ while if $\varphi$ 
has also at least $N$ vanishing moments, then $K_{\varphi,\psi}(x,s) \sim s^{-(N+1)}$ when $s \gg 1$.
In that respect, for each $x$, $K_{\varphi,\psi}(x,s)$ is maximum around $s = 1$.
Moreover, if $\psi$ is also supported by $[-1/2,1/2]$, then $K(x,s) = 0$ if $|x|\geq s$. 
In the sequel, for the sake of simplicity and to avoid cumbersome considerations about 
the tails of $K(x,s)$ for large and small $s$, we will suppose that the kernel $K(x,s)$ is 
non vanishing only in the time-scale interval $s \in [\kappa s,1]$, $x \in [-s,s]$ for some $\kappa < 1$.  Under this assumption, the wavelet transform of $X(t)$ can be approximated as:
\begin{equation}
\label{eq:wt-approx}
W(x,a) \simeq  \left\{
\begin{array}{ll}
  & \!\!\! \int\limits_{
  	\kappa a}^{\min(T,a)}   s^{H-2} ds \int\limits_{x-a}^{x+a}   e^{\omega_{s}(b)}  K_{\varphi,\psi}\left(\frac{x-b}{s},\frac{a}{s} \right)  db, \; \; \mbox{if} \; \kappa a \leq T \\
  &  0 \; \; \mbox{otherwise} 
\end{array}
\right.
\end{equation}

Let us remark that in the case one chooses the analyzing wavelet $\psi(t) = \delta^{(1)}(t)$ with 
\begin{equation} 
\label{eq:def_delta1}
\delta^{(1)}(t) = \delta(t+1)-\delta(t), 
\end{equation}
where $\delta(t) dt$ 
stands for the Dirac distribution, then $W(t,a)$ is nothing
but the increment of $X(t)$ at scale $a$ \cite{MuBaAr93PhysRevE}:
\begin{equation}
\label{eq:poor_man_wavelet}
W(t,a) = \delta_a X(t) = X(t+a)-X(t)
\end{equation}
Despite this "poor man's wavelet" does not possess the requested regularity and oscillating 
properties (in particular approximation \eqref{eq:wt-approx} is not supposed to hold), in the sequel, we will often consider that the increments statistics can be deduced from the 
wavelet transform statistics as a particular case (see e.g. \cite{MuBaAr93PhysRevE} for 
a discussion on this specific topic).
Along the same line, if $\psi(t)= \delta^{(2)}(t)$,
with
\begin{equation}
\label{eq:def_delta2}
\delta^{(2)}(t) = \delta(t+2)+\delta(t)-2 \delta(t+1)
\end{equation}
then $W(t,a)$ corresponds to the increments of second order of $X(t)$: $W(t,a) = X(t+2a)+X(t)-2X(t+a)$.
Hereafter, we will refer to increments of first or second order when it is necessary to distinguish
these two specific types of wavelet transforms.


Let us define the spectrum of structure function scaling exponents
\begin{equation}
\label{eq:def_zeta}
   \zeta_q = qH - \phi(q)
\end{equation}
and consider its Legendre transform:
\begin{equation}
\label{eq:singspectr-wc}
  F(h) = 1+\inf_q(qh-\zeta_q) \; .
\end{equation}
Let us suppose that $\exists \eta > 0$ such that $F(h) < 0$ for all $0 \leq z < \eta$.
In Appendix \ref{App:conv_C0}, we use expression \eqref{eq:wt-approx} to show that, for any $L>0$,  almost surely, 
the paths of $X(t)$ have a uniform Lipschitz regularity $\alpha$ on $[0,L]$ 
for all $0 < \alpha < h_{\min}$ with
$$
 h_{\min} = \arg \max \{F(h) < 0\}_{h < H-\phi'(0)} \; .
$$  
This result about the almost sure regularity of the paths  $X(t)$ extends to continuous $\cal W$-cascades
a similar property proven by Arneodo et al. in the case of discrete $\cal W$-cascades \cite{ArnBacMuz98}.
For example, in the log-normal case, $\phi(q)$ is provided in Eq. \eqref{eq:phiq_ln} and therefore $h_0 = H-\phi'(0) = H+\frac{\lambda^2}{2}$. Since $F(h)= 1-\frac{(h-h_0)^2}{2 \lambda^2}$, we obtain, provided $H > \sqrt{2 \lambda^2} -\frac{\lambda^2}{2}$, i.e., $h_{\min} = h_0-\sqrt{2 \lambda^2} > 0$.

Let us notice that, unlike discrete $\cal W$-cascades, the weighting 
process $s^H e^{\omega_\ell(t)}$ involved in the continuous version
is always positive. Nevertheless, let us remark that wavelet analysis is also helpful to show that, in the construction \eqref{eq:def_X}, instead of choosing the (positive) weights $s^{H} e^{\omega_s(b)}$, one could equivalently build $X(t)$ from a zero mean weighting process that has the same law as $W(b,s)$, the wavelet transform of $X(t)$. Indeed, let us consider the 
following process: 
\begin{equation}
\label{wt-recons}
   X'(t) = \lim_{\ell \to 0} \lim_{R \to \infty} \int_{\ell}^{R} ds \; s^{-2} \! \int_{-\infty}^{+\infty}   W(b,s) \varphi \left(\frac{t-b}{s} \right)  db  
\end{equation}
where $W(b,s)$ is the wavelet transform of $X(t)$ as defined in \eqref{eq:wtX}.
According to the wavelet point-wise reconstruction formula (e.g. as introduced in \cite{HT91}), 
if the analyzing wavelet 
$\psi$ satisfies some mild conditions as respect to the synthetizing wavelet $\varphi$, one expects that $X'(t)=X(t)$ at every point of continuity of $X(t)$. It thus results that
\eqref{wt-recons} provides a whole family of versions of 
continuous $\cal W$-cascades contruction were the weights are chosen to
be a zero mean process as simply obtained from any wavelet transform of the original $X(t)$ function.

\section{Self-similarity and scaling properties}
\label{sec:scaling}
In this section we study the scaling and self-similarity properties of $X(t)$
as defined in Eqs. \eqref{eq:def_X} and \eqref{eq:lim_X}.

\subsection{Stochastic self-similarity of the wavelet transform. Multifractal Scaling}
\label{sec:wc-cascades_ss}
Let us first point out that, from the construction 
of $\omega_{\ell}$ as recalled in Appendix \ref{app:cont_casc},
Eq. \eqref{charf} can be naturally generalized as:
\begin{equation}
\label{charf_2}
\EE{e^{\sum_{m=1}^q i p_m \omega_{\ell_m}(x_m)}}  = e^{\sum_{j=1}^q \sum_{k=1}^{j}
	\alpha(j,k) \rho_{\max(\ell_k,\ell_j)}(x_k-x_j)}
\end{equation}
for any sequence $0, \ell_1,\ldots,\ell_q < T$. From the expression
\eqref{rhoexact} of $\rho_\ell(t)$, 
it results that Eq. \eqref{eq:ss-omega} can be extended as equality in 
law for processes of both space and scale variables: one has, $\forall r < 1$, $0 < \ell \leq T$ and  $u \in [0,T]$,
\begin{equation}
\label{eq:ss-omega2}
\omega_{r \ell}(s u)  \stackrel{\cal L'}{=}  \Omega_r + \omega_\ell(u)
\end{equation}
where where $\stackrel{\cal L'}{=}$ means that the two processes have the same 
finite dimensional distributions of any order as processes in the half-plane
$(\ell,u)$ and $\Omega_r$ is a random variable of same law as $\omega_{rT}(u)$ independent of the process $\omega_\ell(t)$.

Let $r < 1$ and $a \ll T$. From Eq. \eqref{eq:wt-approx}, the rescaled version
of the wavelet transform of $X(t)$ reads:
\begin{eqnarray*}
W(rx,ra) & \simeq & \int\limits_{
	\kappa r a}^{ra}   s^{H-2} ds \int\limits_{rx-ra}^{rx+ra}   e^{\omega_{s}(b)} K_{\varphi,\psi}\left(\frac{rx-b}{s},\frac{ra}{s} \right)  db  \\
         & = & r^{H} \int\limits_{
         	\kappa a}^{a}   s^{H-2} ds \int\limits_{x-a}^{x+a}   e^{\omega_{rs}(rb)} K_{\varphi,\psi}\left(\frac{x-b}{s},\frac{a}{s} \right)  db  \; .
\end{eqnarray*}
Thanks to Eq. \eqref{eq:ss-omega2}, we can thus establish the self-similarity of the 
wavelet transform of $X(t)$:
\begin{equation}
\label{eq:wt-ss}
   W(rx,ra) \stackrel{{\cal L'}}{=}  r^{H} e^{\Omega_r} W(x,a) \; .
\end{equation}

\begin{center}
	\begin{figure}[h]
		\vspace{-0.5cm}
		\includegraphics[width=0.6\textwidth]{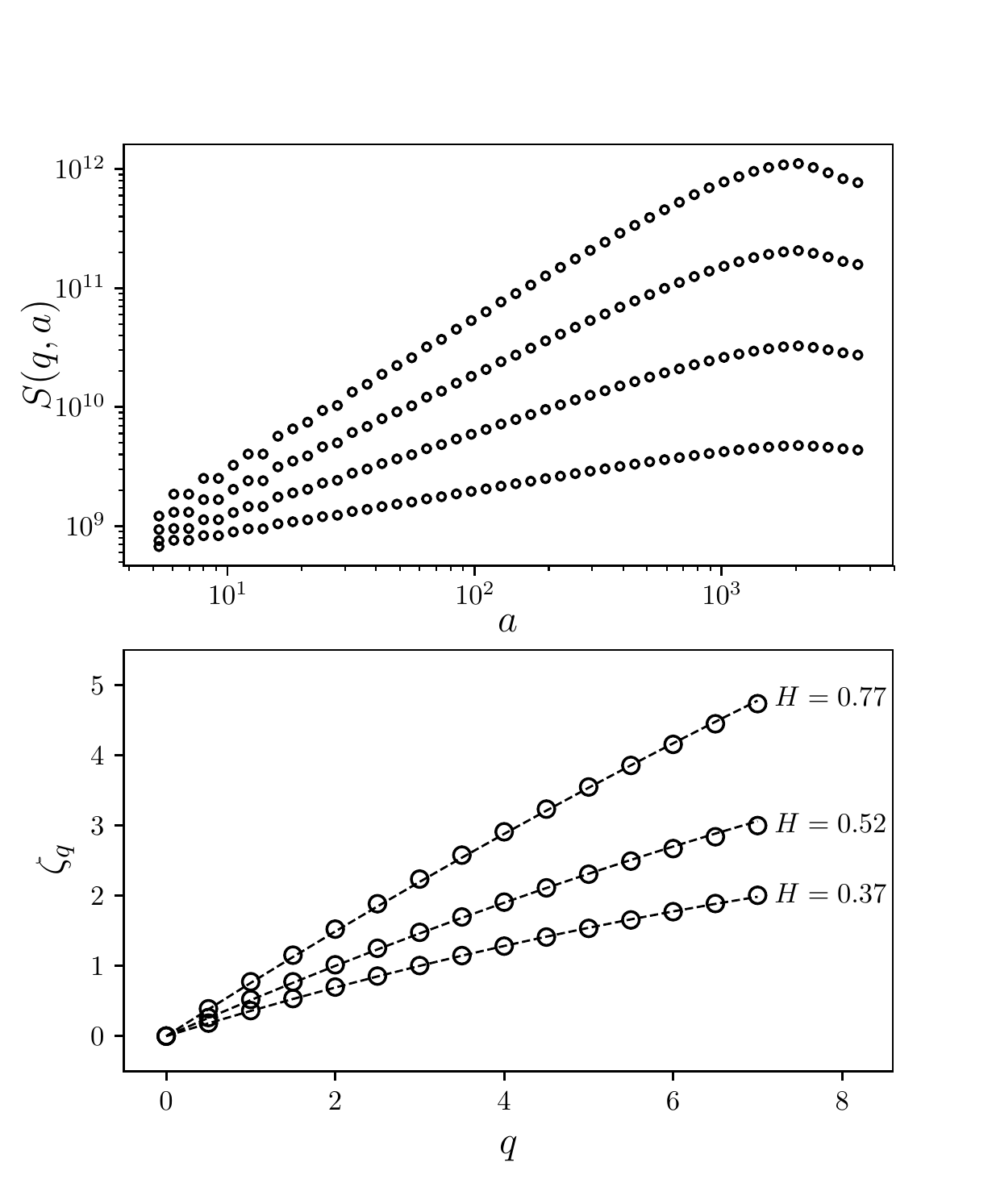}
		\caption{Power-law scaling of structure functions of $X(t)$. In the top panel
			the structure function $S(q,a)$ of $X(t,a)$ (with $H=0.36$) are plotted in double logarithmic scale for $q=1,2,3,4$. The scaling behavior holds up 
			to $a \simeq 2. 10^3$ which corresponds to the integral scale $T$. In the bottom panel, the estimated $\zeta_q$ ($\circ$) are compared to the expected log-normal expressions as given by Eq. \eqref{eq:zetaln} (dashed lines) for $H=0.36$, $0.52$ and $0.77$ and $\lambda^2 = 0.025$.}
		\label{fig_sf}
	\end{figure}
\end{center}

From Eq. \eqref{eq:wt-ss}, the definition of $\Omega_r$ , Eqs. \eqref{eq:def_phi} and \eqref{eq:def_zeta}
it results that, for all $a \leq T$, the wavelet structure functions are characterized by the 
multifractal scaling:
\begin{equation}
\label{eq:wt-sf}
  S(q,a) = \EE{|W(x,a)|^q} = C_q \left(\frac{a}{T}\right)^{qH-\phi(q)} = C_q a^{\zeta_q} \; ,
\end{equation}
where $C_q = \EE{|W(x,T)|^q}$.
$X(t)$ has therefore multifractal properties in the sense
that the absolute moments of its wavelet transform behave has a power-law
with a non-linear concave multifractal spectrum $\zeta_q$ \cite{MuzBacArn91,MuBaAr93PhysRevE}.
In particular, by considering Eq. \eqref{eq:poor_man_wavelet}, we deduce that 
the moments of absolute increments behave has power-laws with the multifractal 
spectrum $\zeta_q$:
\begin{equation}
\label{eq:wt-sf}
\EE{|X(t+a)-X(t)|^q} = C_q a^{\zeta_q} \; .
\end{equation}

This result is illustrated in Fig. \ref{fig_sf} where the estimated structure functions
\begin{equation}
\label{eq:def_sf}
  S(q,a) = \sum_k |X(k+a)-X(k)|^q
\end{equation}
are computed on realizations 
of log-normal versions of $X(t)$ with $H=0.36$, $0.51$ and $0.76$ respectively and 
$\lambda^2 = 0.025$. In each case, the integral scale has been set to $T= 2048$ and the overall sample length corresponds to 16 integral scales. In the top panel of Fig. \ref{fig_sf} we have plotted $\log_{10}S(q,a)$ as a function of $\log_{10} a$ in the case $H=0.36$. We see that a power-law behavior extends from the smallest scale up to the integral scale. The estimated $\zeta_q$ spectrum, as obtained from a linear fit of these
log-log plots, are reported in the bottom panel (symbols $\circ$). In the three cases, we obtain, within a good precision, the expected scaling exponents:
\begin{equation}
\label{eq:zetaln}
 \zeta_q = \left(H + \frac{\lambda^2}{2}\right) q - \frac{\lambda^2}{2}q^2
\end{equation}
represented by the dashed lines.
Let us notice that in the case $H = 0.5125$, one has $\zeta_2 = 2$ 
and therefore the successive increments of $X(t)$ are uncorrelated as in the 
Brownian motion (see next Section). When $H=0.36$ (precisely $H = 0.3585$), one has $\zeta_3 = 1$ as expected for the increments of the longitudinal velocity in fully deve\-loped turbulence \cite{Fri95}.
\begin{center}
	\begin{figure}[t]
		\includegraphics[width=1\textwidth]{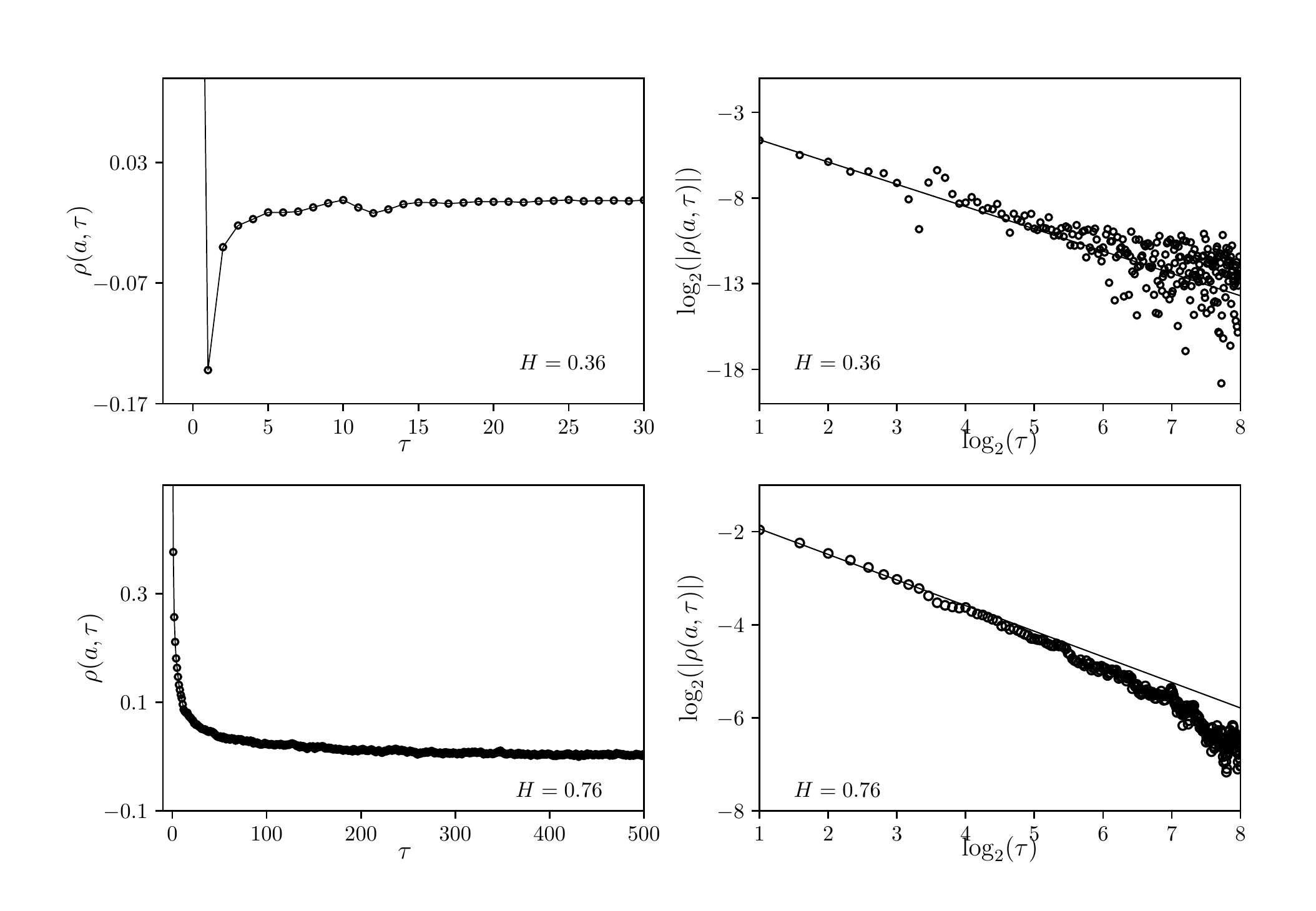}
		\caption{Estimated correlation function of the increments of $X(t)$ for 
			$H = 0.36$ (top panels) and $H=0.76$ (bottom panels). In both cases $\omega_\ell(t)$ is a Gaussian process of intermittency coefficient $\lambda^2 = 0.025$. The synthetizing wavelet $\varphi(t)$ is a "mexican hat" wavelet, $g^{(2)}(t)$ defined in Eq. \eqref{eq:gauss_deriv}. As expected when $H$ is small enough increments are anti-correlated
			while for $H$ large enough. The expected power-law behavior \eqref{eq:inc_cf} are represented by the solid lines in left panels. }
		\label{fig_corr}
	\end{figure}
\end{center}
It is well known that the spectrum of scaling exponents
$\zeta_q$ can be, for a large class of functions, related
to the singularity spectrum, i.e., the fractal (Haussdorf) dimension of the sets of iso-H\"older regularity \cite{ParFri85,Jaf97a,Jaf97b,BM04}. This
is the so-called multifractal formalism.
At this stage, it is tempting to conjecture that the result proven by Barral and Seuret for random wavelet series \cite{BarSeu05} is also valid in the 
framework introduced here and that the multifractal formalism holds for continuous wavelet cascades. In our context, the Barral-Seuret result would say that $D(h)$ the  singularity spectrum of $X(t)$, 
can be simply obtained from $D_M(\alpha)$, the singularity spectrum of the underlying log-infinitely divisible cascade $dM_t = \lim_{\ell \to 0} e^{\omega_{\ell}(t)} dt$.
More precisely, if $D_M(\alpha)$ is the singularity
spectrum of the continuous cascade $M(t)$ as provided 
by the multifractal formalism \cite{BM04}, i.e., in the range where $1+\min_q (q\alpha -\phi(q)) >0 $, one has 
$$
D_M(\alpha) = 1+\inf_q(q(\alpha-1)+\phi(q)) 
$$
and thus, from definition \eqref{eq:singspectr-wc},
\begin{equation}
\label{eq:d=F}
D_M(\alpha) = F(\alpha+H-1) \; .
\end{equation}

Then the analog of Barral-Seuret result (Theorem 1.1 of \cite{BarSeu05}) result would assert that
the singularity spectrum $D(h)$ of $\cal W$-cascades is 
provided by $D(h) = D_M(h+1-H)$ which according to 
\eqref{eq:d=F}, would simply mean that the singularity spectrum of $X(t)$ is $D(h) = F(h)$
in the range where $F(h)$, as given by \eqref{eq:singspectr-wc} is positive. Since $F(h) = 1+\inf_q(qh-\zeta_q)$ is the Legendre transform of the spectrum of scaling exponents of wavelet transform structure functions, this would imply that 
the multifractal formalism holds for continuous $\cal W$-cascades.

\subsection{Increment correlation functions and magnitude covariance}
\label{sec:wc-cascades_corr}
In this section, we study the behavior of various correlation functions associated with the increments (or wavelet coefficients) or the powers of their absolute values.

Let us first define the increment correlation function
\begin{equation}
\label{eq:def_incf}
\rho (a,\tau) = \Cov{\delta_a X(t)}{\delta_a X(t+\tau)} = \EE{\delta_a X(t) \delta_a X(t+\tau)}
\end{equation}
It is easy to show that, when $a \ll \tau$, one has \cite{RamBaMu18}
$$ 
  \rho(\tau) \simeq  \frac{\partial }{\partial \tau} \tau^{-1} \EE{\delta_\tau X(t)^2}
$$ 
and therefore, from Eq. \eqref{eq:wt-sf}, one has:
\begin{equation}
\label{eq:inc_cf}
  \rho(\tau) \simeq A \tau^{\zeta_2-2} \; .
\end{equation} 

Let us remark that if $\zeta_2 < 1$, then the prefactor $A$ is negative and 
the increments are anti-correlated while if $\zeta_2 > 1$, the correlations
are positive. This is reminiscent of the correlations of a fractional Brownian motion \cite{ManVan68,Taqqu} increments where $\zeta_2 = 2H$ and for which the value $H = 1/2$  separates the regimes of negatively and positively correlated fluctuations.
The behavior \eqref{eq:inc_cf} is illustrated empirically using two samples of $X(t)$
with respectively $H=0.36$ and $H = 0.76$ while $\lambda^2 =0.025$ and $T=2048$ in both 
cases. The empirical increments correlation function have been estimated at scale $\tau = 1$ from samples of length 16 integral scales. As expected, one can observe, in the right top and right bottom panels, a power-law behavior in both cases 
and that $\zeta_2-2$ provides a good fit (solid lines). In the left panels, we see that the process with 
$H = 0.36$, corresponding to $\zeta_2 = 0.695$ is characterized by anti-correlated increments while the one with $H=0.76$, corresponding to $\zeta_2 = 1.495$, has positively
correlated increments. 

The behavior correlation function of the absolute power of the increments can
be inferred from the self-similarity properties of the wavelet transform $W(x,a)$.
The same kind of scaling argument as previously used to establish the scaling
of structure functions can be used. Indeed, from Eq. \eqref{eq:wt-ss}, one has, for $a \leq \tau \leq T-\kappa a$:
\begin{equation}
 \EE{|W(x_0,ra_0)|^q|W(x_0+r\tau_0,ra_0)|^p} =  r^{\zeta_{q+p}} \EE{|W(x_0,a_0)|^q|W(x_0+\tau_0,a_0)|^p} 
\end{equation}
Let us now choose $\varepsilon \ll 1$ and set $T' = \frac{T}{1+\varepsilon}$.
Let us consider $r a_0 = a$, $r \tau_0 = \tau$ and 
$a = \varepsilon \tau$.
The previous equation can be rewritten as:
\begin{equation}
\EE{|W(x_0,a)|^q|W(x_0+\tau,a)|^p} = \left(\frac{\tau}{T'} \right)^{\zeta_{q+p}} \EE{|W(x_0,\varepsilon T')|^q|W(x_0+T',\varepsilon T')|^p} 
\end{equation}

If one considers $\varepsilon \to 0$ (i.e. one chooses $a$ very small), then $T' \simeq T$ and $W(x_0,\varepsilon T')$ becomes 
independent from $W(x_0+T',\varepsilon T')$ and one has:
\begin{equation}
 \EE{|W(x_0,\varepsilon T')|^q|W(x_0+T',\varepsilon T')|^p} \simeq 
 \EE{|W(x_0,\varepsilon T')|^q}\EE{|W(x_0+T',\varepsilon T')|^p} \sim K_{p,q}(a) \left(\frac{\tau}{T}\right)^{-\zeta_q-\zeta_q}
\end{equation}
where $K_{p,q}(a)$ is a constant that depends on $a$.
Given the scaling \eqref{eq:wt-sf}, this entails, for a fixed small value of $a$:
\begin{equation}
\label{eq:corr-scaling}
\EE{|W(x_0,a)|^q|W(x_0+\tau,a)|^p}  \sim K_{p,q}(a) \left(\frac{\tau}{T}\right)^{\zeta_{p+q}-\zeta_q-\zeta_p}
\end{equation}
showing that correlation functions of the powers of the wavelet transform absolute value, behave, at a given scale, as a power-law as a function of the time
lag $\tau$ with a scaling exponent $\zeta_{p+q}-\zeta_p-\zeta_q$.
It is important to notice that this exponent does not depend on $H$ and is only provided by the $\phi(q)$ the non-linear part of $\zeta_q$.
\begin{center}
	\begin{figure}[h]
		\includegraphics[width=0.8\textwidth]{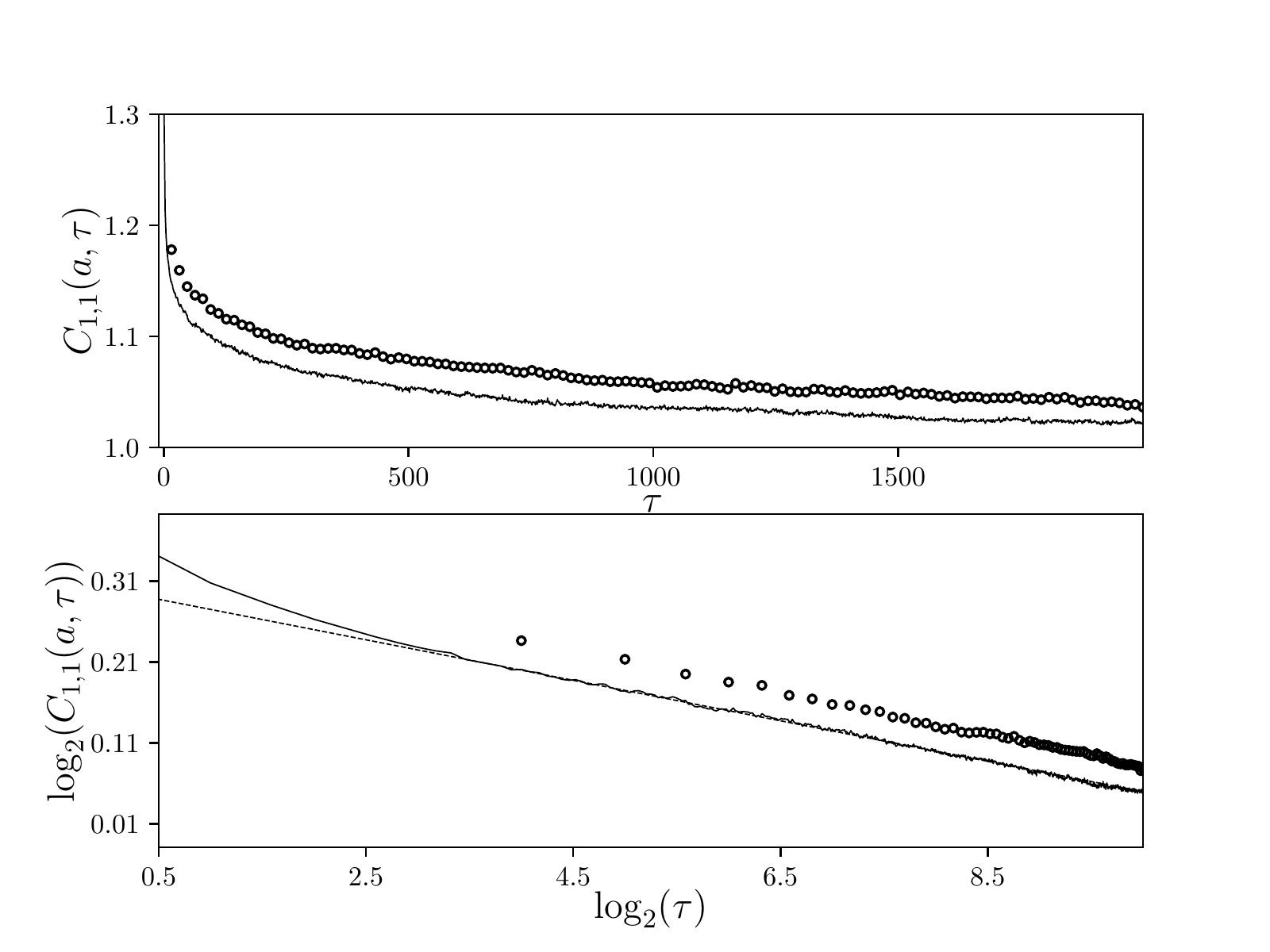}
		\caption{Estimated correlation function of the absolute 
			increments of $X_{LN}(t)$ for $H = 0.37$ ($\circ$) and $H=0.77$ (solid lines). As expected they both behave as $\tau^{-\lambda^2}$ independently of $H$ (represented by the dashed line in the bottom panel).}
		\label{fig_corrabs}
	\end{figure}
\end{center}
This behavior is illustrated in Fig. \ref{fig_corrabs} where we have plotted the estimation of absolute increments (i.e. in the case $p=q=1$) for the two log-normal processes with $H=0.36$ and $H=0.77$ considered previously. As expected both behave, independently of $H$, as $\tau^{\zeta_{2}-2\zeta_1}$ that corresponds, according to Eq.
\eqref{eq:zetaln} to the power-law $\tau^{-\lambda^2}$ (represented by the dashed line in the bottom panel).

Eq. \eqref{eq:corr-scaling} can be used to compute the behavior 
of magnitude covariance as defined in \cite{ArnBaMaMu98,MuDeBa00}. Indeed, since:
\begin{equation}
 \Cov{\ln|W(x_0,a)|}{\ln|W(x_0+\tau,a)|} =  \frac{\partial_q \partial_p \ln \EE{|W(x_0,a)|^q|W(x_0+\tau,a)|^p}}{\partial p \partial q}  \bigg\rvert_{q,p=0} \;,
\end{equation}
we obtain the well known logarithmic magnitude covariance for multifractal
processes \cite{ArnBaMaMu98,MuDeBa00}:
\begin{equation}
\Cov{\ln|W(x_0,a)|}{\ln|W(x_0+\tau,a)|} \simeq \zeta''(0) \ln\left( \frac{\tau}{T} \right) = -\lambda^2 \ln\left( \frac{\tau}{T} \right)
\end{equation}
where we defined the intermittency coefficient as $\lambda^2 = -\zeta''(0)$.

\section{Skewness and leverage effect: applications to turbulence and stock market data}
\label{sec:cw-cascades_skew}

\subsection{Skewness of increment pdf at all scales}
Because of the self-similarity relationship \eqref{eq:wt-ss}, one can 
also expects a scaling of odd moments of the wavelet transform, i.e.,
$\forall k \in \mathbb{N}$, $\forall a \leq T$,
\begin{equation}
\label{eq:sf_odd}
 S^\star(2k+1,a) = \EE{W(x,a)^{2k+1}} = V_{2k+1}  \; a^{\zeta_{2k+1}}
\end{equation}
This shows in particular that generically, the skewness 
$$
\frac{ \EE{W(x,a)^{3}}}{ \EE{W(x,a)^{2}}^{3/2}} \sim a^{\zeta_3-\frac{3}{2} \zeta_2 }
$$
increases (in absolute value) when $a \to 0$. 
The computation of the constant $V_{2k+1}$ for a given waveform $\varphi$ is tedious
and can be only written as an intricate multiple integral but it can 
be shown it is non-zero unless $\varphi$ satisfies very specific conditions (see Fig. \ref{fig_sk}).
This shows that our approach, unlike constructions bases
on classical random measures, generically 
leads to skewed multifractal processes with a skewness that, like
the flatness, increases as one goes from coarse to fine scales.
\begin{center}
	\begin{figure}[h]
		\includegraphics[width=0.8\textwidth]{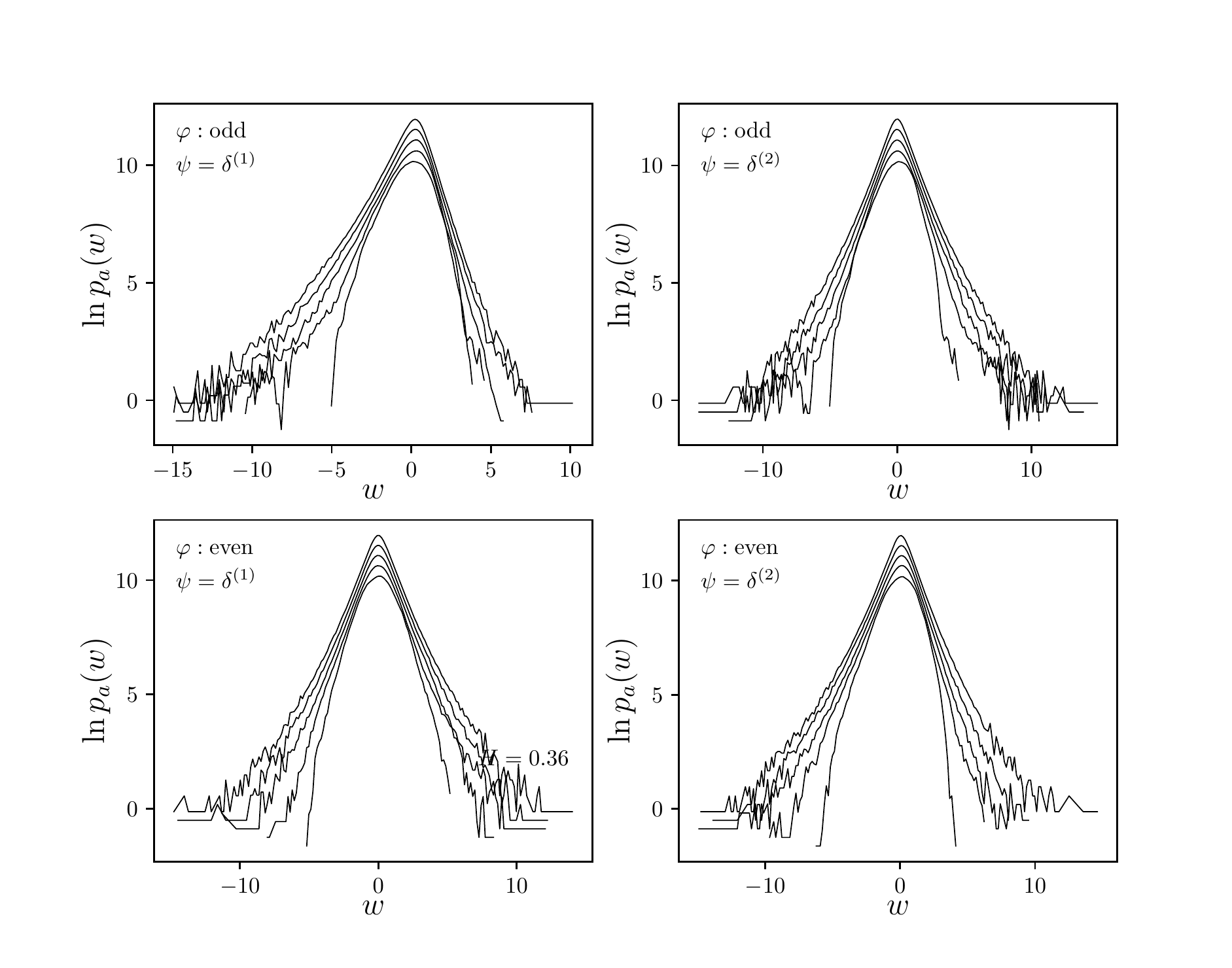}
		\caption{Estimated ``standardized" probability density function (pdf) of the ``wavelet coefficients" at scale $a$, $p_a(w)$. At each scale, $a=T$,$T \over 8$,$T \over 64$,$T \over 512$, $T \over 4096$,
			where $T$ stands for the integral scale, we have displayed $\ln p_a(w)$ up to an additive constant for the sake of clarity: pdf at small scales are displayed above pdf at coarser scales. On the top left and right panels the considered $\cal W$-cascade $X(t)$ is a log-normal process with a synthetizing wavelet $\varphi$ that is a odd function ($g^{(1)}(t)$, the first derivative of the Gaussian function). We can check that, due to the intermittency, the flatness increases from large to small scales. On the left panel the analyzing wavelet is odd since it is $\psi(t)= \delta^{(1)}(t)$ (the one that corresponds to the increments of $X(t)$) while on the right panel one uses $\psi(t)=\delta^{(2)}(t)$ and thus considers the second order increments. As expected, one clearly observes, like in turbulence, skewed distributions of increments at all scales. The second order increments are distributed with a symmetric law. On the bottom panels, the opposite effect is observed since the process $X(t)$ as been constructed using a symmetric wavelet, namely $g^{(2)}(t)$, the second order derivative of the Gaussian function.}
		\label{fig_pdf}
	\end{figure}
\end{center}
\begin{center}
	\begin{figure}[h]
		\vspace{-1.5cm}
		\includegraphics[width=0.8\textwidth]{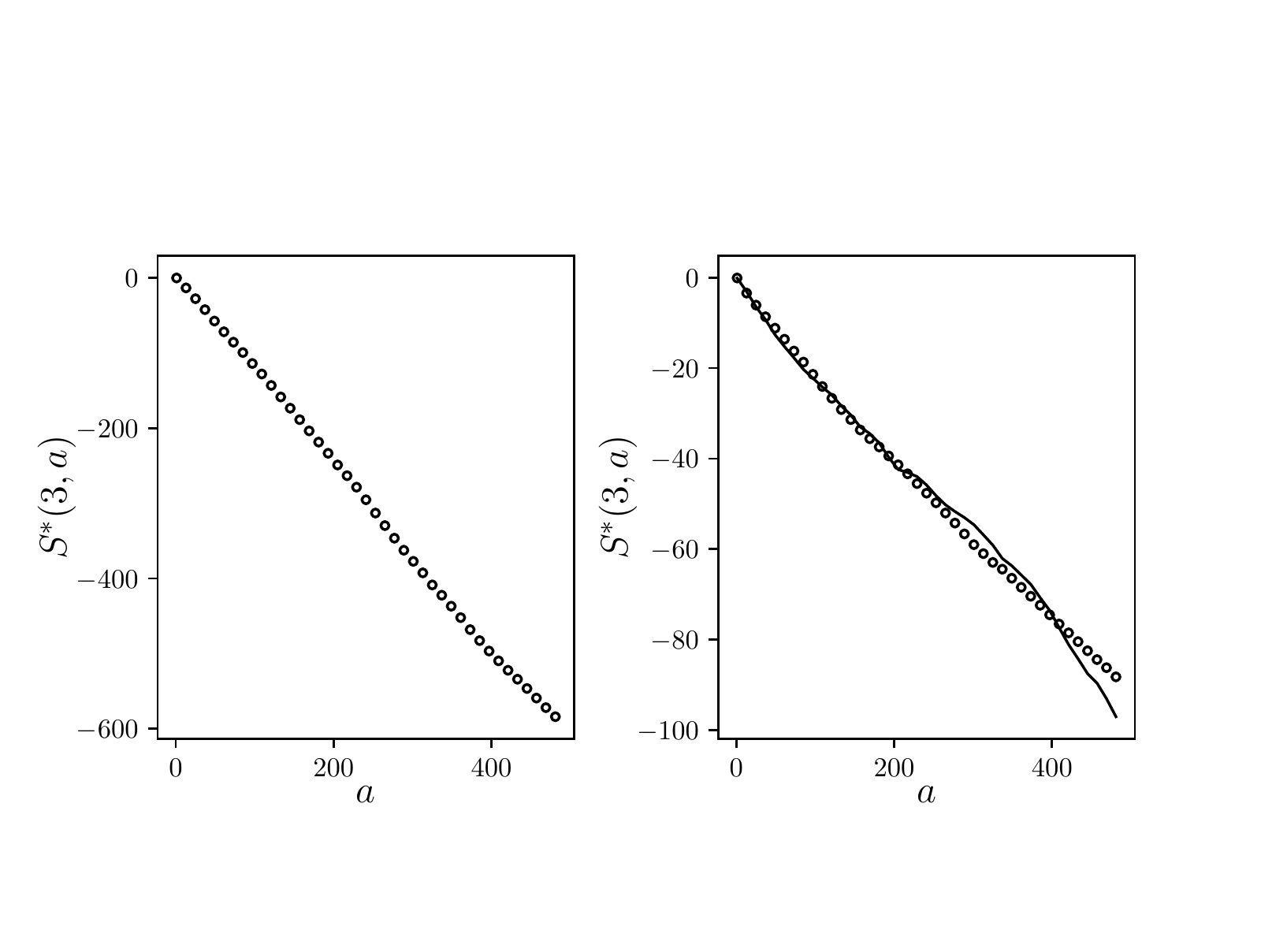}
		\vspace{-1cm}
		\caption{Signed third order structure function $S^\ast(3,a)$ as a function of the scale $a$. On the left
			panel the mean order 3 moment of the increments is computed on $X(t)$ corresponding to the anti-symmetric
			synthetizing wavelet $g^{(1)}(t)$ (Eq. \eqref{eq:gauss_deriv}). This confirms the (negative) skewness already illustrated in Fig. \ref{fig_pdf} and that, as expected for turbulence, $S^\ast(3,a)$  behaves as a linear function. In the right panel, $X(t)$ is built using the non-symmetric wavelet depicted in Fig. \ref{fig_wavelets}.
			The third order structure function function behaves as a linear function and skewness can be observed for both first ($\circ$) and second order increments (solid line).}
		\label{fig_sk}
	\end{figure}
\end{center}
\vspace{-1.5cm}
The scaling relationship \eqref{eq:sf_odd} indicates that 
the skewness will be zero
at all scales if it vanishes at the largest scale $T$. It is noteworthy that this may be the case if the synthetizing wavelet $\varphi$ in Eq. \eqref{eq:def_X} is an even or odd function. Indeed, if the wavelet $\varphi(t)$ is a symmetric function, because $\omega_{\ell}(t) \stackrel{{\cal L}}{=} \omega_{\ell}(-t)$, we see from definition \eqref{eq:def_X} that $X_\ell(t)$ is invariant by 
time reversal, i.e., one has $X_\ell(-t) \stackrel{{\cal L}}{=} X_\ell(t)$. It thus results that if the analyzing wavelet an odd function 
(as e.g. for the increments $\delta_{\tau} X(t)$) the wavelet transform of $X(t)$ will have a symmetric
law implying that all odd moments are zero. In order to observe some skewness in the increment law, it is thus necessary to consider non-symmetric synthetizing wavelets $\varphi$. Along the same line, if the synthetizing wavelet $\varphi(t)$ is anti-symmetric, 
then $X(t)$ is odd by time reversal, i.e.,
$ X_\ell(-t)  \stackrel{{\cal L}}{=} -X_\ell(t)$
and if the analyzing wavelet is even (as e.g. when one computes 
the second order increments of $X(t)$, $\delta^{(2)}_\tau X(t) = X(t+2 \tau)+X(t)-2 X(\tau)$), the wavelet transform will have 
a symmetric law. This means that if a skewness is observed in both first and second order increments, the wavelet $\varphi(t)$ is neither an even nor an odd function.
\begin{center}
	\begin{figure}[h]
		\includegraphics[width=1\textwidth]{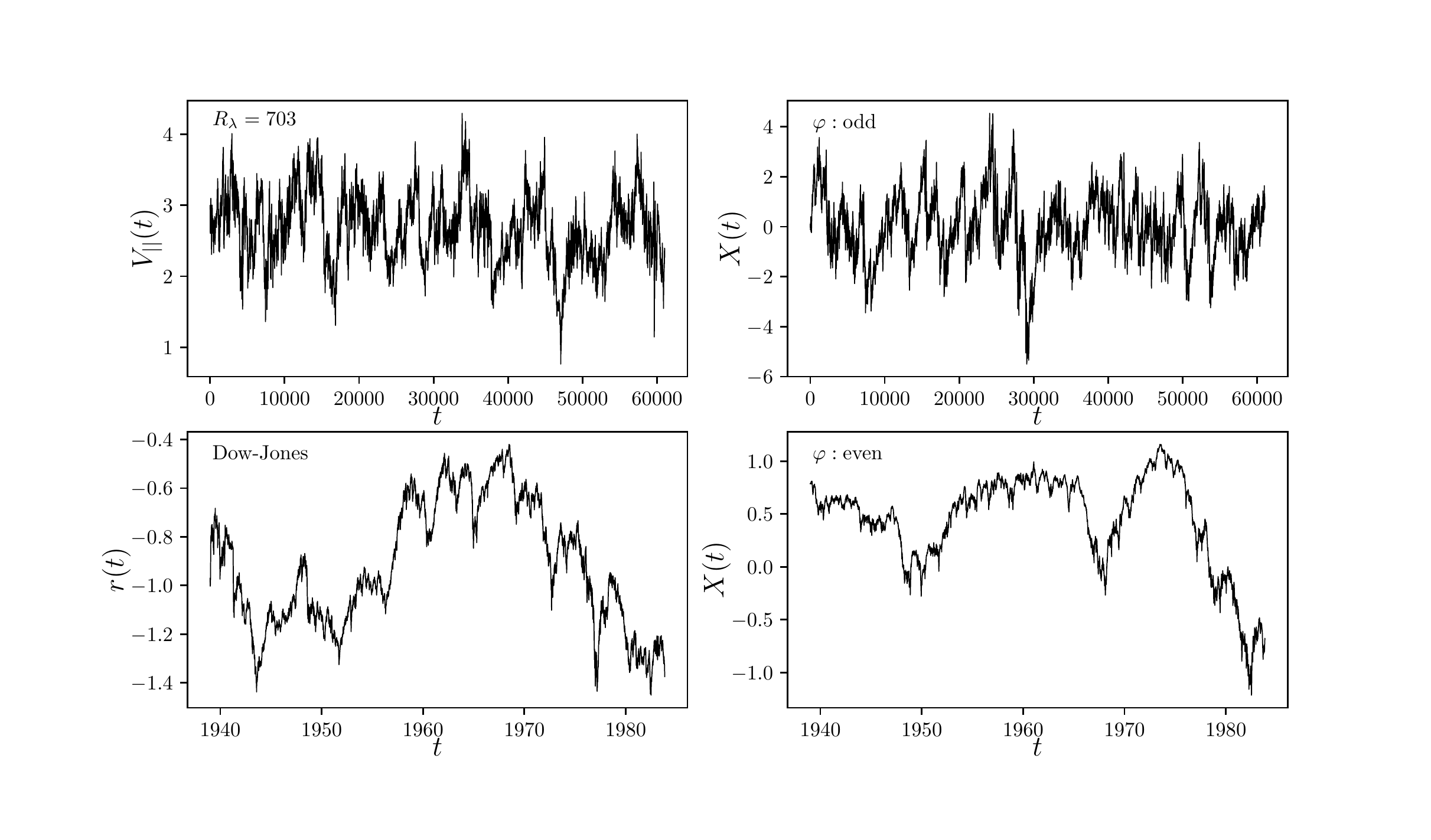}
		\vspace*{-1cm}
		\caption{Paths of turbulence velocity field and stock market data as compared to their $\cal CW$-cascade models. In the top left panel are represented the longitudinal velocity field space variations (in $m/s$) obtained in a high Reynolds number turbulence experiment ($R_\lambda \simeq 700$) realized by Castaing et al. \cite{CasHe94}. The top right panel corresponds to  the cascade model obtained by fitting the turbulence multiscaling but also the skewness of law of first order increments.	In the left bottom panel is plotted the detrended daily variations of the logarithm of the Dow-Jones index over the more than 4 decades (from 1939 to 1985). The right bottom panel represents its $\cal CW$-cascade model where we mainly account for the skewness of second order increments. We can see that the "ramp" like behavior of turbulence data and "arch" like patterns of stock market data are remarkably well reproduced by ${\cal CW}$-cascades.}
		\label{fig_paths2}
	\end{figure}
\end{center}

These behaviors are illustrated in Fig. \ref{fig_pdf} where whe have displayed the probability density functions (pdf)
of the first and second increments at different scales 
of two versions of log-normal continuous wavelet cascades $X(t)$ that are respectively antisymmetric and symmetric by time reversal. In the first case, we used $\varphi = g^{(1)}$  while in the second case $\varphi = g^{(2)}$ (see Eq. \eqref{eq:gauss_deriv}). In both cases, we chose $\omega_{\ell}$ as a log-normal cascade with $\lambda^2 = 0.025$, an integral scale $T = 2^{12}$ and $H=0.355$. In that respect, according to \eqref{eq:zetaln}, $\zeta_3 = 1$ and $X(t)$ was designed to mimic the main features of velocity records in experiments of fully developed turbulence. All the estimations have been performed on a sample of length $128$ integral scales. All the probability density functions reported in Fig. \ref{fig_pdf} are standardized, i.e., represent the distribution of increments normalized by their root-mean-square. The are displayed in semi-log scales and shifted so that large scale distribution are below fine scale ones. In that way, one can clearly observe the
intermittency as an increasing of the flatness from large to small scales. We can check in the top-right (resp. bottom-left) panel of Fig. \ref{fig_pdf} that if $\varphi$ is odd (resp. even), the second (resp. first) order increments are symmetrically distributed. We can see the in top-left panel that, at all scales the increments of the anti-symmetric version of $X(t)$ are negatively skewed. The shapes of these skewed pdf, with an increasing flatness are strikingly similar to distribution of longitudinal velocity increments in fully developed turbulence \cite{CaGaHo90,CasHe94}. 
In the symmetric version of $X(t)$, the skewness is only observed on its second order increments (bottom-right panel).

The fact that an anti-symmetric synthezing wavelet allows one to reproduce the observed skewness behavior of the velocity field
in turbulence is directly illustrated in left panel of Fig. \ref{fig_sk}: we have plotted, in linear scale,
the signed third order structure function $S^\star(3,a)$ as a function of the scale $a$ estimated for a log-normal $\cal CW$-cascade
calibrated precisely to model the spatial fluctuations of the longitudinal velocity in turbulence 
(i.e., one sets $\zeta_3 = 1$ and $\lambda^2 = 0.025$). 
It can be checked that for such a process we have $ S^\star(3,a) = -K a$. 
In full analogy with turbulence, one can
wonder if $K$ could not be chosen such that, as in Kolmogorov theory, $K = \frac{4}{5} \varepsilon$ where $\varepsilon$ is the mean dissipation rate $\varepsilon \sim \EE{ a^{-1} \nu \int_0^a dt  (\frac{\partial}{\partial t} X(t))^2} \;$  \cite{Fri95}.
Since $X(t)$ is not differentiable, in order for this to be meaningful, we show,
in Appendix \ref{App:diss_anomaly},  that the so-called ``dissipative anomaly'' \cite{EySree06} property of the velocity field 
can be reproduced within our framework: if one chooses a ``viscosity" $\nu(\ell)$ such that $\nu(\ell) \sim \ell^{4/3-\lambda^2}$, then:
\begin{equation}
\label{eq:diss_an}
 \lim_{\ell \to 0} \nu(\ell) \; \EE{ \left(\frac{\partial X_\ell(t)}{\partial t }\right)^2 } = \varepsilon \; ,
\end{equation}
for some $0 < \epsilon < \infty$.
Let us point out that it is likely that, from the definition Eq. \eqref{eq:def_X}, one could establish the following 
generalization of Eq. \eqref{eq:diss_an}:
$$
\nu(\ell) \left( \frac{\partial X_\ell}{\partial t}  \right)^2 dt \underset{\ell \rightarrow 0}{\longrightarrow}  \varepsilon(dt)
$$
where $\varepsilon(dt)$ represents a singular measure corresponding to the multifractal dissipation \cite{menesree91} in the limit of vanishing viscosity and the convergence being interpreted in a weak sense.

Since the (negative) skewness of turbulence fields appears mainly on first order increments (or on odd analyzing wavelets),
as one can see in the top panels of Fig. \ref{fig_paths2}, it can be directly visualized as ``ramp" like (i.e. a slow increase followed by a rapid fall) patterns on the 
velocity profile and its model. This behavior is very different from the "arch" like shapes that can be associated with the  negative skewness of the second order increments. Such a feature is obviously present in the fluctuations of market prices as illustrated in the example of the Dow-Jones index the bottom left panel of Fig. \ref{fig_paths2}. In the right panels, we plotted sample paths of the corresponding log-normal models for turbulence ($H=0.355$, $\lambda^2 =0.025$ and $\varphi = g^{(1)}$) and stock market data 
($H=0.51$, $\lambda^2=0.025$ and $\varphi = g^{(2)}$).

A close inspection of the Dow-Jones evolution in the bottom-left panel of Fig. \ref{fig_paths2} reveals 
that this process has not only skewed second order increments by also display skewed first order increments.
Indeed, it is well known that financial return variations are characterized by 
rapid large drops that are followed by slower upward moves. It thus results that, in order to model
the dynamics of market prices, a $\cal CW$-cascade should involve a wavelet that is neither odd nor even.
In the left panel of Fig. \ref{fig_sk}, we have computed the behavior of $S^\star(3,a)$ as a function of $a$ 
for both first and second order increments when the wavelet is the non-symmetric wavelet displayed 
in Fig. \ref{fig_wavelets}(c) (the other parameters are those chosen for turbulence).
We see that both types of increments are characterized by a skewness of comparable magnitude since one observes third order moments a behaves as similar linear functions of the scale.
As discussed below, asymmetric synthetizing wavelets allow one to account for another feature observed on stock market data, namely the leverage effect.

\subsection{Leverage effect}
The fact that the sythetizing wavelet $\varphi(t)$ has no particular symmetry can be reflected
by different statistical quantities.  As discussed notably by Pommeau \cite{Pom82}, there exists a wide variety of correlation functions
that allow one to reveal the lack of time reversal symmetry of a process $X(t)$.
The ``leverage'' function is a particular example of such a measure. It consists in computing the correlation between
the increments at some scale and their "amplitude" (e.g. their absolute value and their squared value) after or before some time lag $\tau$ \cite{BouMatPot01,PocBou02,BaDuMu12}:
\begin{equation}
\label{eq:def_lev}
 {\cal L}_q(\tau) = Z_{q,\ell}^{-1} \EE{\delta_\ell X(t) |\delta_{\ell} X(t+\tau)|^q}
\end{equation}
where $Z_{q,\ell}^{-1}$ is a properly chosen normalization constant: for example, in \cite{BouMatPot01}, 
the authors studied ${\cal L}_2$ with $Z_{2,\ell} = \EE{\delta X_\ell^2}^2$.
In general one will consider
the behavior in the small scale regime, i.e., the limit $\ell \to 0$ with a wel chosen normalization.
The so-called ``leverage effect'' is a property that has been empirically observed on stock prices and stock market indices 
indicating that past returns are (negatively) 
correlated with future volatility (i.e. ${\cal L}_q(\tau) < 0$ for $\tau \geq 0$) while the reverse is not true (${\cal L}_q(\tau) \simeq 0$ for $\tau < 0$). 
In that context "returns" means increments of log-price while "volatility" stands for squared or absolute returns. The leverage effect can be intuitively explained by the fact that, after a large price drop, some kind of panic takes place with a large uncertainty
and thus a large volatility results. The symmetric situation, i.e a large upward 
price move, does not induce any crisis and therefore does not trigger any noticeable volatility variation.
\begin{center}
	\begin{figure}[h]
		\includegraphics[width=0.7 \textwidth]{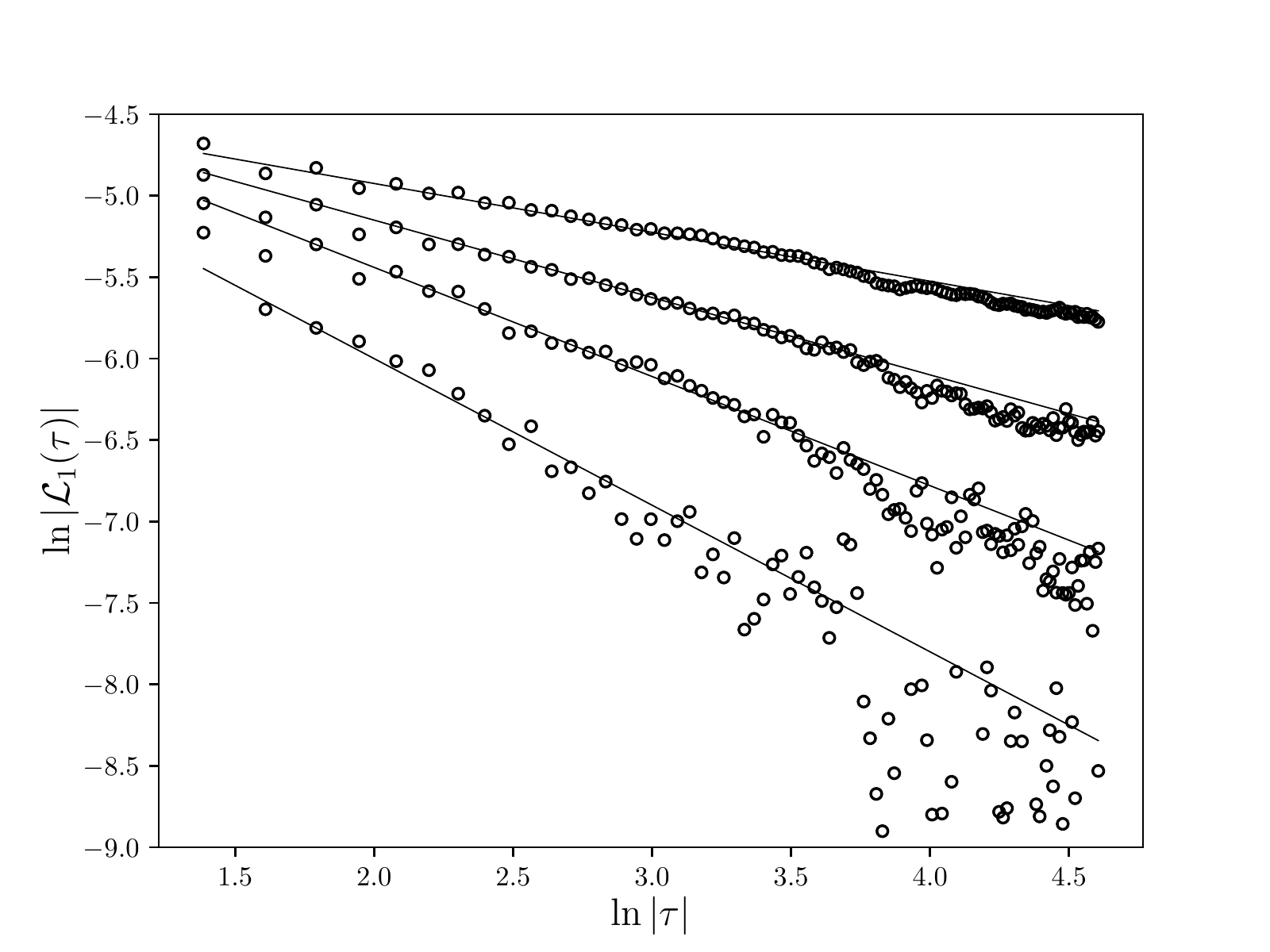}
		\caption{Scaling of the leverage function. Estimated curves $|{\cal L}_1(\tau)|$ are plotted as a function of $|\tau|$ is double logarithmic scale (symbols ($\circ$)). The estimations have been performed on samples of lenght $L = 2^{17}$ of log-normal ${\cal CW}$-cascades with $\lambda^2 = 0.025$, $T = 2^{10}$ and, from top to bottom, $H=0.7, 0.51, 0.33$ and $0.1$.
			The solid lines represent the corresponding theoretical power-laws of exponent $H-1-\phi(2)$.}
		\label{fig_Levloglog}
	\end{figure}
\end{center}
\begin{center}
	\begin{figure}[h]
		\includegraphics[width=0.6 \textwidth]{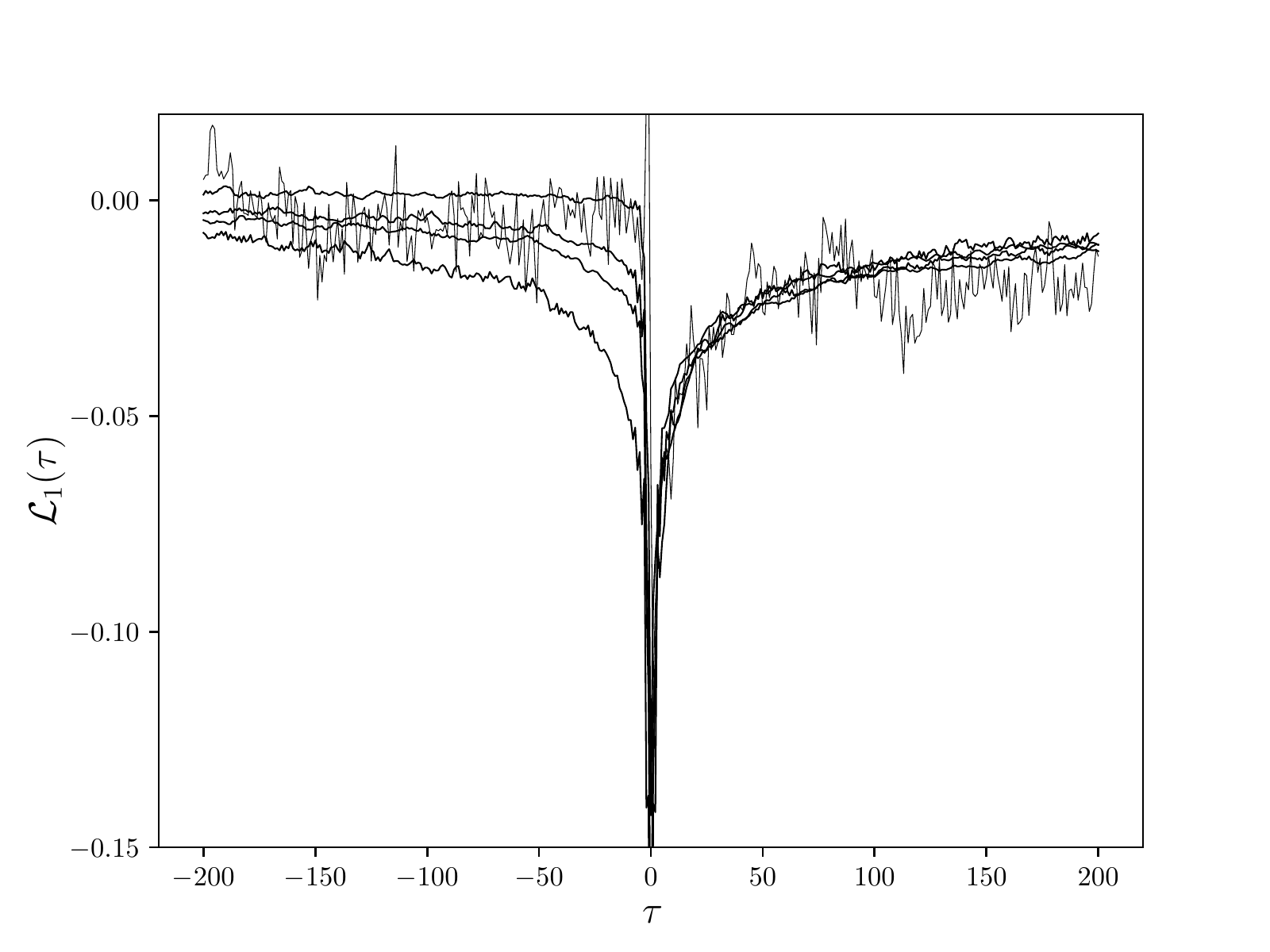}
		\caption{Leverage functions ${\cal L}_1(\tau)$ estimated for 
			a log-normal $\cal{CW}$-cascade with $H=0.515$ and $\lambda^2 = 0.025$ (solid lines). The various curves correspond to different asymmetry factors $\alpha$ in the synthetizing wavelet defined in Eq. \eqref{eq:asym_wavelet}, namely $\alpha = \frac{1}{8}$, $\frac{1}{3}$, $\frac{1}{2}$ and $1$.  One sees that the behavior at lags $\tau >0$ does not depend on $\alpha$ whereas the leverage function at negative lags becomes smaller and smaller as the asymmetry increases. For comparison purpose, the estimated leverage function from daily Dow-Jones 
			index data are also reported (grey curve).}
		\label{fig_LevFunction}
	\end{figure}
\end{center}

Different attempts to account for this effect within the standard class of econometric models have been proposed \cite{BouMatPot01,PerMas03}. Since the class of Multifractal Random Walk as described in Sec. \ref{sec:c-casc} remarkably accounts for many 
of ``stylized facts" of asset fluctuations, some authors considered different variants of these models 
that break the time reversal symmetry 
by introducing specific correlations between cascade and noise terms \cite{PocBou02,BaDuMu12}. But as mentionned in the introductory
section, such approaches cannot lead to well defined continuous time limits unless the noise has long-range correlations \cite{Poc03,BaDuMu12,Che17}.
Since continuous $\cal W$-cascades are generically not invariant by time reversal one expects that it should be possible 
to account for the leverage effect by a specific choice of the synthetizing wavelet $\varphi$.
According to the definitions \eqref{eq:def_lev} and \eqref{eq:def_X}, ${\cal L}_q(\tau)$ can be expressed as 
an intricate integral.
In Appendix \ref{App:Leverage}, we show, using some heuristic approximations, that, when $q=1$, the leverage function 
behaves as:
\begin{eqnarray*}
 {\cal L}_1(\tau) & \underset{\tau \gg \ell}{\sim} & C_{+} |\tau|^{H-1-\phi(2)} \\
 {\cal L}_1(\tau) & \underset{\tau \ll -\ell}{\sim} & C_{-} |\tau|^{H-1-\phi(2)}
\end{eqnarray*}
where both scaling laws are valid in the range $|\tau| \ll T$.
We have shown that the prefactors $C_{+}$ and $C_{-}$  depend on the synthetizing wavelet $\varphi$ as:
\begin{eqnarray*}
	C_{+}  & = &  -\int_{0}^{+\infty} \! \! s^{H-1} ds \int  \! du  \varphi \left(u \right) C_s'(su+1) \\
	C_{-}  & = &  -\int_{0}^{+\infty} \! \! s^{H-1} ds \int  \! du  \varphi \left(u \right) C_s'(su-1) \\
\end{eqnarray*}
with $C_s(z) = \EE{e^{\omega_s(t)+\omega_s(t+z)}}$.
This scaling law is illustrated in Fig. \ref{fig_Levloglog} where
we have checked, using estimations from simulated data, it holds for different values of $H$ in the case of a log-normal cascade built with an anti-symmetric wavelet.
Notice that from the previous expression, $C_-$ and $C_+$ are 
not necessarily equal and the leverage effect could therefore be observed when the leverage ratio:
\begin{equation}
\label{eq:leverageratio}
 \kappa = \frac{C_+}{C_-}
\end{equation}
is very large (i.e. $\kappa \gg 1$).
\begin{center}
	\begin{figure}[h]
		\includegraphics[width=0.7 \textwidth]{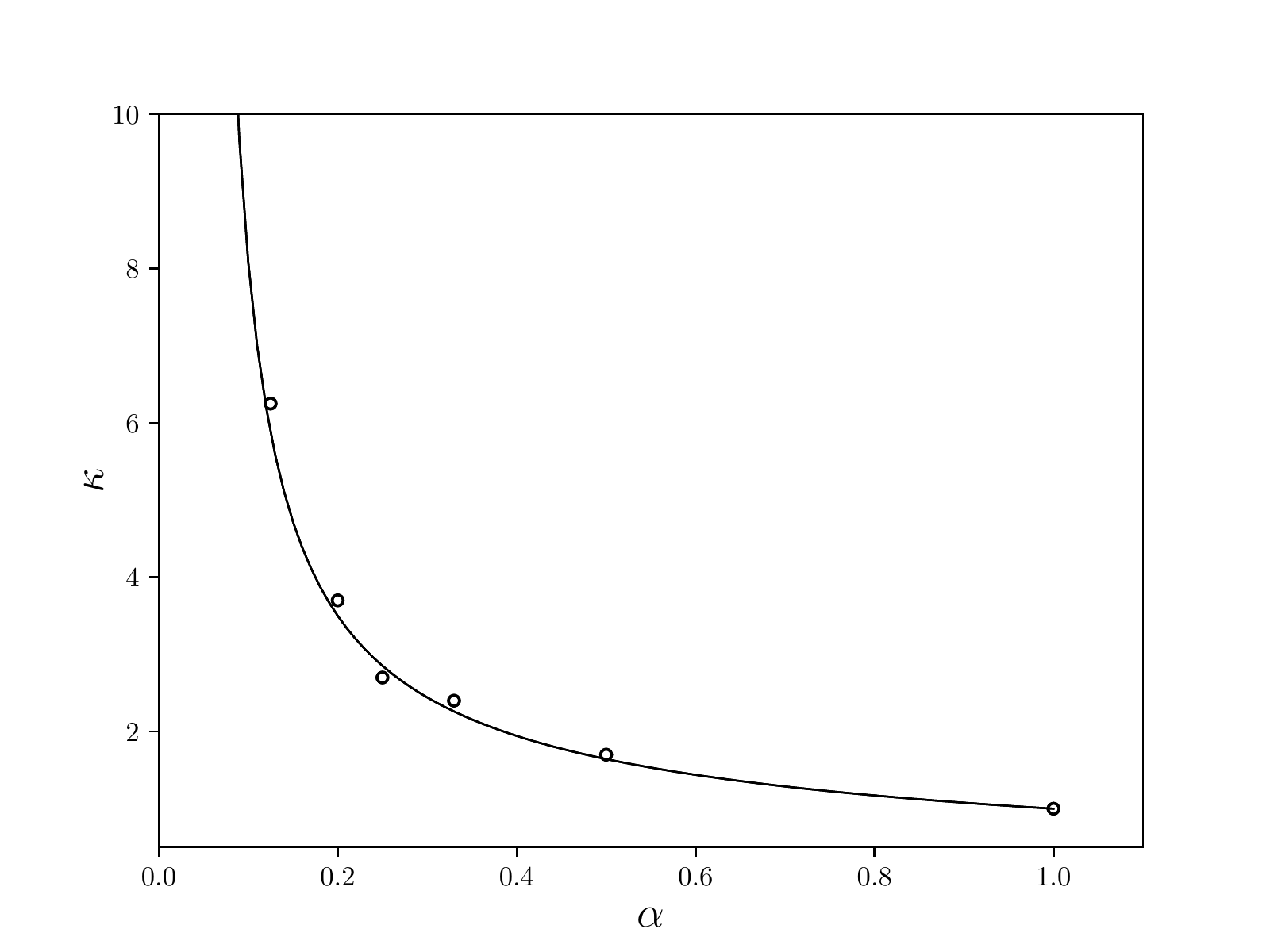}
		\caption{Evolution of the leverage ratio $\kappa$ (Eq. \eqref{eq:leverageratio}) as a function of the coefficient $\alpha$ characterizing the asymmetry of the synthezing wavelets defined in \eqref{eq:asym_wavelet}. The numerical integrations have been performed in the log-normal case with $\lambda^2 = 0.025$ and $H = 1/2+\lambda^2$. The symbols ($\circ$) represent the value
			of $\kappa$ as estimated from the empirical leverage function obtained from simulated data of length 128 $T$ with $T = 2^{13}$.}
		\label{fig_ratio}
	\end{figure}
\end{center}
For illustration purpose and in order to handle reasonable expressions, let us consider the ``simple'' 
case when  $\varphi(t)$ is a piece-wise constant function:
\begin{equation}
\label{eq:asym_wavelet}
\varphi(t) = \alpha^{-1}\mathbb{I}_{[0,\alpha)}(t)-\mathbb{I}_{[-1,0)}(t)
\end{equation}
where $0<\alpha<1$ controls the wavelet asymmetry. 

In Fig. \ref{fig_LevFunction} we have plotted the leverage function 
estimated from numerical simulations of 
log-normal ${\cal CW}$-cascades with a synthezing wavelet corresponding to Eq. \eqref{eq:asym_wavelet} with different asymmetry factors $\alpha= 1/8,1/3,1/2$ and $1$. We chose $T = 2^{13}$, $\lambda^2 = 0.025$ and $H=0.51$ so that the increments are uncorrelated ($\zeta_2 = 1$). We first clearly see that the behavior for positive lags appears to not depend on $\alpha$ while the situation is very different for negative lags: as $\alpha$ decreases one sees that the leverage function lessens more and more. One goes from a symmetric leverage function in the case $\alpha=1$ to a situation where it is almost zero at negative lags. This latter case is very interesting to reproduce the empirical features observed on stock market: for comparion we have plotted the leverage function estimated from the Dow-Jones index daily returns over a period extending from 1939 to 2018 (grey line in Fig. \ref{fig_LevFunction}).  
These empirical findings show that the leverage 
ratio strongly depends $\alpha$ and becomes arbitrary large 
when $\alpha$ is small. Even if it is possible to show that
the constants $C_\pm$ are bounded in the case when $\varphi$ is defined by Eq. \eqref{eq:asym_wavelet}, their exact value can however be hardly obtained or approximated under a closed form. We then estimated them by means of a numerical integration.
The so computed leverage ratios are reported in Fig. \ref{fig_ratio}.
In the case when $\alpha = 1$, we naturally recover the 
fact that the leverage effect is not present while we observe that
$\kappa$ strongly increases for small $\alpha$.
Notice that our numerical estimations confirm that both $C_+$ and $C_-$ are negative in the range $0.1 \leq \alpha \leq 1$. We also observed that $C_+$ is almost independent of $\alpha$ while $C_{-}$ becomes arbitrary small as $\alpha \to 0.048$ where it changes its sign.
It thus seems possible, within this model, to obtain a "perfect" leverage effect with an infinite leverage ratio corresponding to a vanishing leverage function in the domain $\tau <0$.
These numerical computations of the leverage ratio have been checked 
using empirical estimation from numerical simulations (symbols $\circ$).

\section{Summary and prospects}
\label{sec:conclusion}
In this paper we have proposed a new way to build random
multifractal functions with stationary increments, exact scaling and self-similarity properties. 
Our model just consists in extending former
$\cal W$-cascades by replacing the framework of orthogonal 
wavelet basis by that of continuous wavelet transform 
and the discrete multiplicative weights by their log-infinitely divisible counterpart, i.e. the process $e^{\omega_{\ell}(t)}$ 
involved in the construction of continuous cascade measures. We have shown that our construction provides almost surely Liphscitz regular paths and studied its self-similarity and scaling properties. As emphasized in Sec.\ref{sec:cw-cascades_skew}, $\cal CW$-cascades are in general skewed multifractal processes and are characterized by
a non-symmetric correlation between increment signs and amplitudes (the so-called "leverage effect").
As far as applications to turbulence are concerned, since our framework can easily reproduce the skewed intermittency phenomenon observed for the velocity fluctuations, together with the dissipative anomaly, it provides undoubtedly 
a promising way to account for many stochastic aspects of fluid dynamics in regimes of fully developed turbulence. 
It that respect, vector and 3-dimensional extensions of $\cal CW$-cascades with the possibility of introducing well
known dynamical properties like incompressibility (as e.g. in \cite{CheRobVar10,ChePeGa16}) could be interesting in order to get a more realistic model.

Beyond applications to specific contexts and the previously mentioned problems, 
a fundamental question is to know which features of the sythetizing wavelet remain observable through the associated continuous cascade. 
Is there a analog of the famous black holes "no-hair theorem" \cite{MTW73} in the present framework ? This "inverse problem" is interesting since we already known that some important properties like 
skewness, the behavior of the leverage function, the prefactor values in scaling relationships may depend on the specific wavelet shape but the issue is to precisely know in what respect and also which properties of $\varphi$ (like e.g. the number of vanishing moments, the moment values,...) can be recovered from empirical data.

Finally, let us mention that, by simply replacing
$e^{\omega_{\ell}(t)}$ by its lacunary version introduced in \cite{MuBai16}, our framework may also allow one to build random functions that are almost everywhere smooth and singulular on random Cantor sets. One could also slightly extend the definition of ${\cal CW}$-cascades in order to build a stationary variant of the class of lacunary wavelet series that possess oscillating singularities defined 
and studied by Arneodo {\em et al.} \cite{ABJM97,ABJM98}. Such processes
are not self-similar but possess a self-similar wavelet transform. 
All these prospects will be considered in a future research.

\appendix

\section{Continuous cascade construction}
\label{app:cont_casc}
The process $\omega_{\ell}(x_0)$ in the definition \eqref{eq:def_M} is constructed as follows: 
On considers the time scale half-plane $(t,s) \in \mathbb{R} \times \mathbb{R^{+ \ast}}$ and 
the natural measure $dm(t,s)= s^{-2} dt ds$ which gives the area of any set ${\cal S}$ as: 
$$
|{\cal S}| = \int_{\cal S} dm(t,s) = \int_{\cal S} s^{-2} dt ds   \;.
$$ 
Within this framework, one considers $dP(t,s)$ a random infinitely
divisible "white noise" (the so-called "independently scattered random measure") such that,
the measure of a given set ${\cal S}$, $P({\cal S}) = \int_{\cal S} dP(t,s)$ is an infinitely 
divisible random variable of characteristic function:
\begin{equation}
\EE{e^{ik P({\cal S})}} = e^{|{\cal S}| \phi(ik)}
\end{equation}
where $\phi(q)$ is the cumulant generating function associated with 
an infinitely divisible law as provided by the celebrated Levy-Khintchine Theorem \cite{Fel71}.
For example, if $\phi(q) = qH - \lambda^2 q^2/2$, $dP(t,s)$ is simply a Gaussian white noise of mean $H s^{-2} dt ds$ and
variance $\lambda^2 s^{-2} dt ds$.

Let us now, as in \cite{MuBa02,BaMu03}, consider $dP(t,s)$ such that $\phi(1) = 0$ and define, for any $T>0$, the cone like domain $\cA_a(x_0)$ as:
\begin{equation}
\label{def:cone}
(x,s) \in \cA_a(x_0)  \Leftrightarrow \left\{s \geq a, \; 2 |x-x_0| \leq \min(s,T) \right\}
\end{equation}

\begin{center}
	\begin{figure}[h]
		\includegraphics[width=0.6 \textwidth]{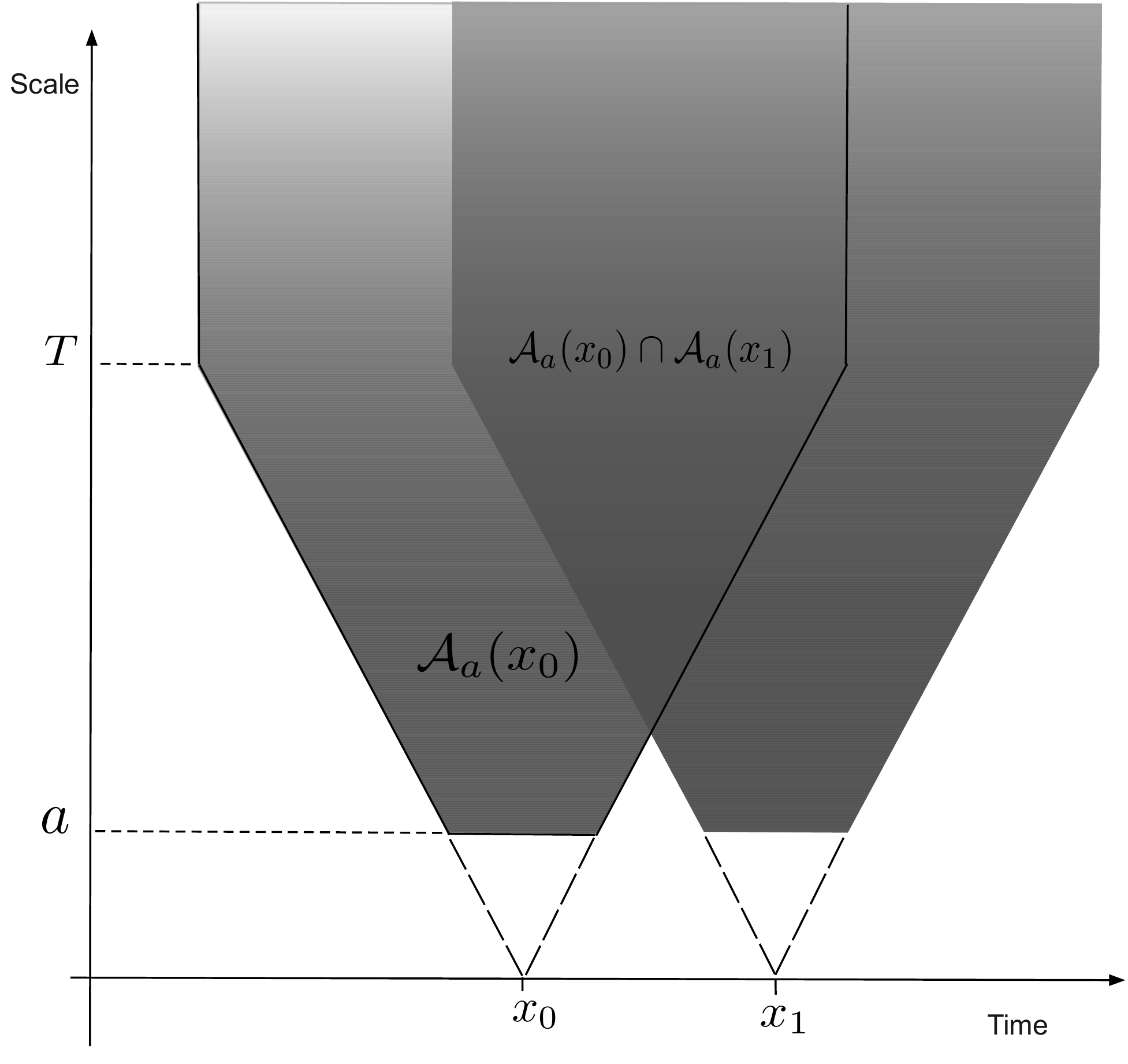}
		\caption{The domains $\cA_a(x_0)$ and $\cA_a(x_1)$ and their intersection.}
		\label{fig1}
	\end{figure}
\end{center}

The shape of $\cA_a(x_0)$ is depicted in Fig. \ref{fig1}. The process $\omega_{\ell}(x)$ is then simply 
defined as:
\begin{equation}
\label{eq:def_omega}
\omega_\ell(x) = P \left[\cA_\ell(x) \right] \; .
\end{equation}
Let $\rho_{\ell}(\tau)$ be 
the area of the set $\cA_\ell(x_0) \cap \cA_\ell(x_0+\tau)$ (see Fig. \ref{fig1}). A direct computation
leads to:
\begin{equation}
\label{rhoexact}
\rho_{\ell}(\tau) =
\left\{
\begin{array}{ll} 
\ln \left(\frac{T}{\ell} \right)+1-\frac{\tau}{\ell} & \mbox{if}~\tau \leq \ell \; , \\ 
\ln \left(\frac{T}{\tau} \right) & \mbox{if}~T \ge  \tau \ge \ell \; , \\
0 & \mbox{if}~ \tau > T \; .
\end{array}
\right. 
\end{equation} 
In \cite{BaMu03} it is shown that the characteristic function of $\omega_{\ell}(x)$, for any $q \in \mathbb{N}^*$, 
$(x_1,x_2,\ldots,x_q) \in \mathbb{R}^q$ with $x_1 \le x_2 \le \ldots \le x_n$ and $(p_1,p_2,\ldots,p_q) \in \mathbb{R}^q$ is given by:
\begin{equation}
\label{charf}
\EE{e^{\sum_{m=1}^q i p_m \omega_{\ell}(x_m)}}  = e^{\sum_{j=1}^q \sum_{k=1}^{j}
	\alpha(j,k) \rho_\ell(x_k-x_j)}
\end{equation}
whare $\alpha(j,k)$ are coefficients 
defined in \cite{BaMu03} that satisfy:
\begin{equation}
\label{rk}
\sum_{j=1}^q \sum_{k=1}^{j}\alpha(j,k) = \phi \left( \sum_{k=1}^q p_k \right)  \; .
\end{equation}


Expression \eqref{charf} entails in particular that $\rho_{\ell}(\tau)$ corresponds to the covariance (when it exists) of $\omega_{\ell}(x)$ and $\omega_{\ell}(x+\tau)$. Notably,
in the case when $dP(t,s)$ is a Gaussian white noise, the above construction of $e^{\omega_{\ell}(t)}$ is an example of the celebrated Kahane Multiplicative 
Chaos measure \cite{kah85,rhova14} and corresponds to the measure originally proposed in the MRW construction \cite{MuDeBa00, BaDeMu01}.

\section{Weak convergence of $X_{2^{-n}}(t)$ in the space of continuous functions.}
\label{App:conv_C0}
Let us show that the series
\begin{equation}
X^{(n)}(t) = X_{2^{-n}} (t) = \int_{2^{-n}}^T \! s^{H-2} ds  \! \int_{-\infty}^{+\infty} \! \! \!   e^{\omega_{s}(b)} \varphi \left(\frac{t-b}{s} \right)  db 
\end{equation}
converges in the weak sense when $n \to \infty$ in the space of continuous functions.

We first need a result that can be found e.g. in \cite{MuBai16} that can be directly deduced 
from the definition of $\omega_s(t)$ in Section \ref{app:cont_casc}: for all $s' \leq s \leq T$, one has: 
$$
 \EE{e^{\omega_s(u)+\omega_{s'}(v)}} = e^{(\phi(2)-2\phi(1))\rho_s(u-v) + \phi(1)(\rho_s(0)+\rho_{s'}(0))}
$$
where $\rho_s(\tau)$ is defined in Eq. \eqref{rhoexact}. From this expression, one deduces:
$$
\EE{e^{\omega_s(u)+\omega_{s'}(v)}} \leq C_1 s^{\phi(1)-\phi(2)}s'^{-\phi(1)}
$$
i.e., assuming \eqref{eq:phi1}, 
\begin{equation}
\label{eq:bb1}
\EE{e^{\omega_s(u)+\omega_s'(v)}} \leq C_1 s^{-\phi(2)}
\end{equation}

Let us first show that all finite dimensional distributions of $X^{(n)}$ converge.
For that purpose it suffices to show that for all $m,n \to 0$
$$
\EE{\left(X^{(n)}(t)-X^{(m)}(t)\right)^2} \to 0
$$
and therefore, assuming $n \leq m$:
$$
\EE{\left(\int_{2^{-m}}^{2^{-n}} \! s^{H-2} ds  \! \int_{-\infty}^{+\infty}   e^{\omega_{s}(b)} \varphi \left(\frac{t-b}{s} \right)  db \right)^2} \to 0 \; .
$$
By permuting expectation and integration, we have to show that
\begin{equation*}
 \iint_{[2^{-m},2^{-n}]} ds_1 ds_2 (s_1 s_2)^{H-2} \iint db_1 db_2 \varphi \left(\frac{t-b_1}{s_1} \right)  \varphi \left(\frac{t-b_2}{s_2}\right) \EE{e^{\omega_{s_1}(b_1)+\omega_{s_2}(b_2)}}  \to 0
\end{equation*}
Since $\varphi$ is bounded and supported by $[-1/2,1/2]$, the previous expression can be bounded 
by :
\begin{equation*}
2 C_2 \int_{2^{-m}}^{2^{-n}} ds_1 \int_{s_1}^{2^{-n}} ds_2 \; (s_1 s_2)^{H-2}  \int_{-s_1/2}^{s_1/2} \int_{-s_2/2}^{s_2/2} db_1 db_2   \EE{e^{\omega_{s_1}(b_1)+\omega_{s_2}(b_2)}} 
\end{equation*}
	
and thus, thanks to \eqref{eq:bb1}
$$
\EE{\left(X^{(n)}(t)-X^{(m)}(t)\right)^2} \leq C_3 2^{-n (2H-\phi(2))} \ .
$$
$X^{(n)}(t)$ is therefore a Cauchy sequence provided 
\begin{equation}
\label{eq:conv1-cond}
  \phi(2) < 2 H \; .
\end{equation}
In order to prove the weak convergence, it remains to establish the tightness of the sequence \cite{Bil68}. Since the sequence $X^{(n)}(t)$ are continuous processes, from \cite{swanson2007}, it suffices to show that
\begin{equation*}
 \sup_n \EE{|X^{(n)}(t)|^\nu} <  \infty \; \mbox{and} \; \sup_n \EE{|X^{(n)}(t)-X^{(n)}(u)|^\beta} \leq C |t-u|^\gamma
\end{equation*}
for some positive $\nu, \beta$ and $\gamma$.
Let us show that both assertions hold for $\nu = \beta = 2$ if one supposes that \eqref{eq:conv1-cond} is satisfied. In that case,
by the same kind of computation as previously, thanks to inequality \eqref{eq:bb1}, one can show that
$\EE{|X^{(n)}(t)|^2} \leq C$ where $C$ does not depend on $n$. Moreover, we have:
\begin{eqnarray*}
	 \EE{(X^{(n)}(t)-X^{(n)}(u))^2} & = & \int_{2^{-n}}^T ds_1 \int_{s_1}^{T} ds_2 (s_1s_2)^{H-2} \iint db_1 db_2 \; \EE{e^{\omega_{s_1}(b_1)+\omega_{s_2}(b_2)}}  \\ & & \left(\varphi \left(\frac{t-b_1}{s_1} \right) - \varphi \left(\frac{u-b_1}{s_1}\right) \right) \left( \varphi \left(\frac{t-b_2}{s_2}\right) -\varphi \left(\frac{u-b_2}{s_2}\right) \right) 
\end{eqnarray*}
Let us choose $\gamma$ such that 
\begin{equation}
\label{cond2}
 0 < 2 \gamma < 2H-\phi(2) \; .
\end{equation}
Then the last expression can be bounded provided $\varphi(t)$ belongs to the uniform H\"older space  $C^\gamma(\mathbb{R})$,  i.e., $\exists K_\gamma < \infty$ such that, $\forall t,u$, $|\varphi(t) -\varphi(u)| \leq K_\gamma |t-u|^\gamma$.
We thus have, using Eq. \eqref{eq:bb1} and condition \eqref{cond2}:
\begin{equation*}
\EE{(X^{(n)}(t)-X^{(n)}(u))^2} \leq   K' |t-u|^{2 \gamma}
\end{equation*}
where 
$$
K'=K_\gamma \int_{0}^T ds_1  \int_{s_1}^{T}   ds_2 s_1^{H-1-\gamma} s_2^{H-1-\phi(2)-\gamma} = K_\gamma \frac{T^{2H-2\gamma-\phi(2)}}{(H-\gamma)(2H-2\gamma-\phi(2))}
$$ 
does not depend on $n$. This ends the proof of the tightness of the sequence and thus 
establishes the weak convergence of the sequence $X^{(n)}$ towards a continuous process $X(t)$.

\section{Almost sure pathwise global regularity of $X(t)$}
\label{App:reg_wavelets}
\begin{center}
	\begin{figure}[h]
		\includegraphics[width=0.8 \textwidth]{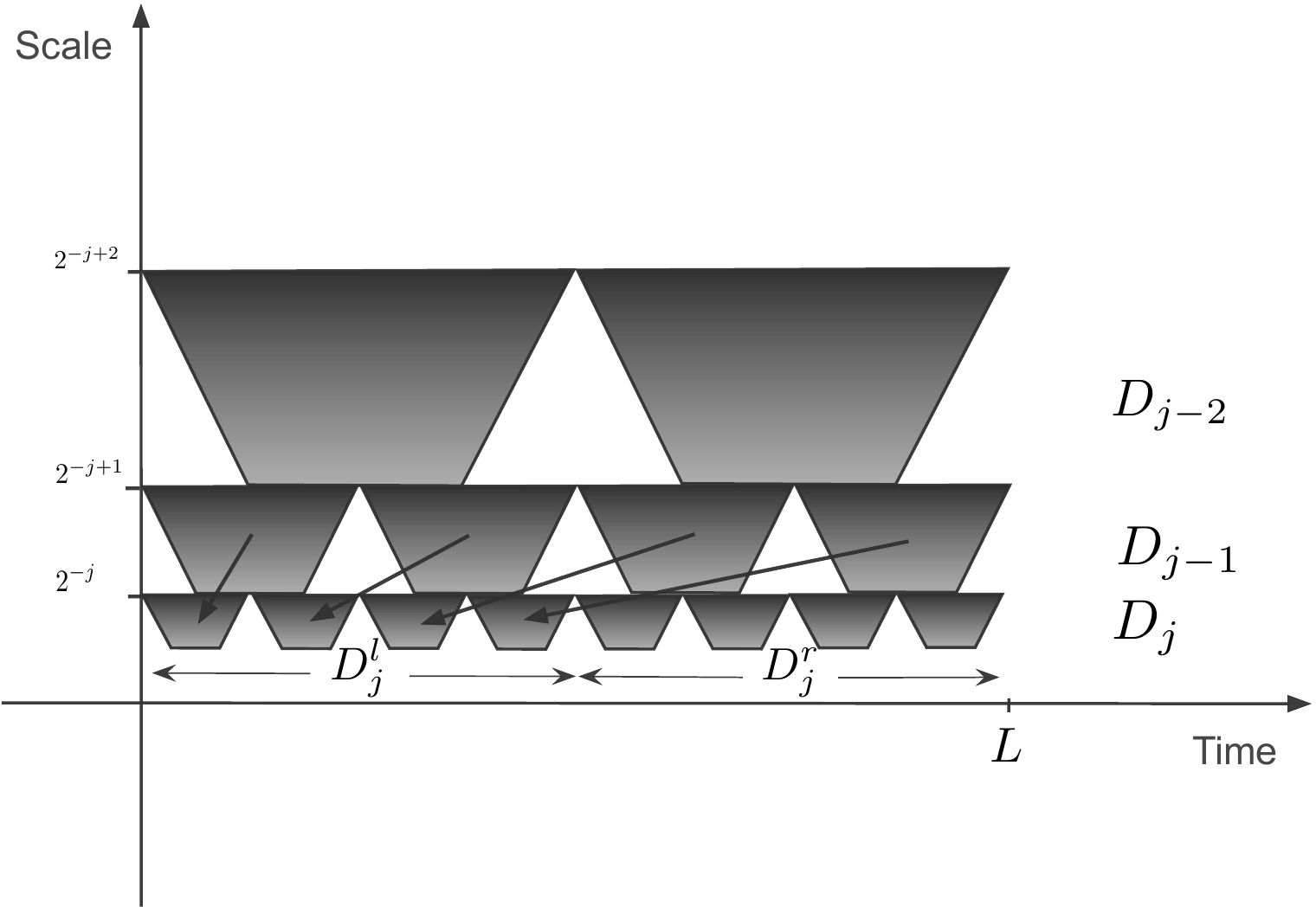}
		\caption{The domains $L_j$ involved in the proof of $L^2$ convergence. Each shaded domain represents a 
		set $D(j,k)$ while $L_j$ is the set of indices $k$ for the $D(j,k)$ that cover the interval $[0,L]$.}
		\label{figApp1}
	\end{figure}
\end{center}

Let us consider $h_{\min}>0$ as defined in Eq. \eqref{eq:def_hmin}.
We then show that, for any $L>0$,  $X(t)$ has a uniform Lipschitz regularity $\alpha$ for all $\alpha < h_{\min}$ on $[0,L]$.
For that purpose, we adapt the proof of Ref. \cite{ArnBacMuz98}
originally proposed for discrete ${\cal W}$-cascades on orthogonal wavelet basis.
Let $\psi_{j,k}(t) = 2^j \psi (2^j t-k)$ be a basis of $L^2([0,L])$ of compactly supported wavelets. For the sake of simplicity, we will not care about boundary wavelets  and we will only consider wavelets  $\psi_{j,k}$ that constitute a basis of $L^2(\mathbb{R})$ which support $S_{j,k}$ is such that $S_{j,k} \cap [0,L] \neq \emptyset$ (see e.g \cite{Mallatbook} for details about wavelet bases on an interval). We refer to $L_j$ the set of indices $k$ that satisfy this property.
We can, without loss of generality, assume that $|S_{j,k}| = 2^{-j}$.
Let $c_{j,k}$ be the wavelet coefficients $X(t)$:
\begin{equation}
\label{eq:dwtX}
c_{j,k} = \int  \psi_{j,k}(t) X(t) dt \; .
\end{equation}
From expression \eqref{eq:wt-approx}, since the kernel $K$ is bounded, $c_{j,k}$ can be controlled as:
\begin{equation}
\label{eq:cwtbound}
|c_{j,k}| \leq K 2^{-jH} \sup_{t,s \in D(j,k)} e^{\omega_{s}(t)}
\end{equation}
where $D(j,k)$ stands for the domain in the time-scale plane $s \in [\kappa 2^{-j},2^{-j}]$, $t \in [2^{-j}k -\frac{s}{2},2^{-j} k +\frac{s}{2}]$ (these domains are depicted as shaded regions in Fig. \ref{figApp1}).

In order control the regularity of $X(t)$, one can use a standard result in wavelet analysis: $X(t)$ 
is uniformly Lipschitz $\alpha < 1$ over $[0,L]$ if and only if there exists an uniform constant $C$ and 
some integer $0 \leq J < \infty$:
that, 
\begin{equation}
\label{eq:Lipcond}
  |c_{j,k}| \leq C 2^{-j \alpha} \;, \;\forall j \geq J, k \in L_j \; .
  \end{equation} 
Let $D_j$ be the time-scale set $D_j$ is defined as (see Fig. \ref{figApp1}):
\begin{equation}
\label{eq:defD_j}
D_j = \bigcup_{k \in L_j} D(j,k)
\end{equation}
and define
\begin{equation}
m_j = \sup_{(t,s) \in D_j} 2^{-jH}  e^{\omega_{s}(t)}
\end{equation}

From \eqref{eq:cwtbound} and \eqref{eq:Lipcond}, we thus have to control the probability that $m_j > 2^{-j \alpha}$ in order to establish global Lipschitz regularity $\alpha$ of $X(t)$.

%

Let us suppose that $L=2^M < T$ where $T$ is the integral scale of $\omega_s(t)$, a proof for an arbitrary value of $L$ can be easily adapted by splitting the interval in small pieces of size smaller than $T$.
Let us now consider that $D_j = D_j^l \cup D_j^r$ where $D_j^{l}$ and $D_j^r$ are 
the union of the $D(j,k)$ for the $2^{M-1}$ values of $k$ corresponding to respectively the first and the second half-part of $L_j$ (see Fig \ref{figApp1}).  
Let $m_j^l = \sup_{(t,s) \in D_j^l}  2^{-jH}  e^{\omega_{s}(t)}$ and  $m_j^r = \sup_{(t,s) \in D_j^r}  2^{-jH}  e^{\omega_{s}(t)}$. 
We have obviously $m_j = \max(m_j^l,m_j^r)$. By stationarity of the process
$\omega_s(t)$, $m_j^r$ and $m_j^l$ have the same law. Moreover, from the construction of $\omega_s(t)$ as the integral
of an infinitely divisible noise over a cone domain, they both can we written as $X^l Y$ and $X^r Y$ where
$X^r$ and $X^l$ have the same law, and $X^l, X^r$ and $Y$ are independent. As shown in \cite{ArnBacMuz98}, it results
that:
\begin{equation}
\label{eq:b3}
 \PP{m_j \leq z} \geq \PP{m_j^l \leq z}^2
\end{equation}
Let us notice that one can map the domain $D_{j-1}$ used to define $m_{j-1}$ 
to the domains $D_{j}^l$ by simply considering the time-scale dilation $s \to s/2$ and $x \to x/2$.
Such a mapping is illustrated by the arrows in Fig. \ref{figApp1}.
Thus, from the self-similarity property of $\omega_s(t)$ (Eq. \eqref{eq:ss-omega}), we have:
\begin{eqnarray*}
\left\{ 2^{-jH} e^{\omega_s(t)} \right\}_{(t,s) \in D_j^l} & = & \left\{ 2^{-jH} e^{\omega_{s/2}(t/2)} \right\}_{(t,s) \in D_{j-1}} \\
		 & \stackrel{\cal L}{=} & \; 2^{-H} e^{\Omega_{1/2}}  \left\{ 2^{-(j-1)H} e^{\omega_{s}(t)} \right\}_{(t,s) \in D_{j-1}}
\end{eqnarray*}
where $\Omega_{1/2}$ is a random variable independent of the process $\omega_s(t)$  such that
$$
\EE{e^{q \Omega_{1/2}}} = 2^{\phi(q)} \; .
$$
This notably implies that:
$$
 \PP{m_j^l \leq z} = \PP{2^{-H} e^{\Omega_{1/2}} m_{j-1} \leq z}
$$
By recurrence we thus obtain:
$$
\PP{m_j \leq z)} \geq \PP{m_0 2^{-jH} e^{\sum_{i=1}^j \Omega_{1/2}^{(i)}}  \leq z}^{2^j}
$$
where $\Omega_{1/2}^{(i)}$ are independent copies of $\Omega_{1/2}$. It thus results that
\begin{eqnarray}
\nonumber
\PP{m_j \geq z} & \leq & 1-\left(1-\PP{m_0 2^{-jH} \; e^{\sum_{i=1}^j \Omega_{1/2}^{(i)}}  \geq z}\right)^{2^j} \\ \label{eq:b4}
                & \leq & 2^j \PP{m_0 2^{-jH} \; e^{\sum_{i=1}^j \Omega_{1/2}^{(i)}}  \geq z}
\end{eqnarray}

Let us now consider $z = 2^{-\alpha j}$ and set $W = 2^{-H} e^{\Omega_{1/2}}$.
Then, provided 
$$-\infty < \alpha < -\EE{\log_2(W)} = H-\phi'(0) \; ,$$
according to Lemma 2 of ref. \cite{ArnBacMuz98} (relying on standard large deviation results),  
for any $\varepsilon >0$, there exists $J_0>0$ such that, $\forall \; j>J_0$
$$
  \PP{m_0 2^{-jH} e^{\sum_{i=1}^j \Omega_{1/2}^{(i)}}  \geq 2^{-\alpha j}} < e^{\varepsilon j} 2^{j F(\alpha)}
$$ 
where $F(\alpha) = 1+\inf_q (q\alpha-\zeta_q)$ with $\zeta_q = qH-\phi(q)$.
Let us suppose that there exists $\eta >0$ such that $F(\alpha) < 0$ in some interval $[0,\eta)$ and 
that 
$$h_0 = H-\phi'(0) >0 \; .$$
Then by choosing $0< \alpha < \eta$ we have, by combining previous inequality and inequality \eqref{eq:b4}:
$$
 \PP{m_j \geq 2^{-\alpha j}} < e^{\varepsilon j} 2^{j (F(\alpha)+1)} 
$$
so one can choose $\epsilon$ small enough such that 
$$
 \sum_j  \PP{m_j \geq 2^{-\alpha j}} < \infty
$$
that means, thanks to Borel-Cantelli Lemma, that almost surely, there exists $J$ such that $m_j < 2^{-\alpha j}$
for $j > J$. We directly deduce that, almost surely, $X(t)$ is a Lipschitz $\alpha$ function.
We can therefore conclude that $X(t)$ is almost surely uniformly Lipschitz $\alpha$
for all $\alpha < h_{\min}$ with
\begin{equation}
\label{eq:def_hmin}
 h_{\min} = \sup \{h, F(h) < 0 \; \mbox{and} \;  h < H-\phi'(0) \}
\end{equation}

\section{Simulation method}
\label{app:simus}
We provide some precisions about the way we performed 
the numerical simulations used in the paper. Let $r >0$ and define
$\varepsilon_{r}[k]$ as an infinitely divisible 
noise defined such that 
$$
 \EE{e^{q \varepsilon_r[k]}} = e^{r \phi(q)}
$$
with $\phi(1) = 0$.
For instance, in the Gaussian case, $\phi(q) = \lambda^2 q(q-1)/2$ and, if $\Delta W_k$ represents an i.i.d. $N(0,\lambda^2)$ Gaussian white noise, then $\varepsilon_r[k] = \sqrt{r} \; \Delta W_k - r \lambda^2/2$.

Let $\theta_a(z)$ be the indicator function of the interval 
$[-a/2,a/2]$ and $\varphi_a(z)= \varphi(z/a)$ be the synthetizing wavelet at scale $a$. Let $\star$ stand for the (fast) discrete convolution operator.
In order to generate a sample of $X_\ell(t)$ over an interval 
at sampling rate $\Delta t$, one considers a discrete version
of Eq. \eqref{eq:def_X}:
\begin{equation}
\label{eq:def_X_d}
  X_\ell(k \Delta t) = \sum_{n=1}^{N} \Delta s_n \; s_n^{H-2}  (\varphi_{s_n} \star e^{\omega_{s_n}})[k \Delta t]
\end{equation}
where $N$ is the number of scales $s_n$ used in the approximation
and $\Delta s_n = s_n-s_{n-1}$.
In practice these scales are chosen as a geometric series, i.e.,  
$s_n = \ell e^{vn}$ where $v$ is such that $s_N = T$.

In order to implement this sum, it is helpful to
remark that, according to definitions \eqref{def:cone} and \eqref{eq:def_omega}, $\omega_{s_n}[k\Delta t]$ can 
be approximated as:
$$
  \omega_{s_n}[k \Delta t] =  (\theta_{s_N} \star \varepsilon_{ s_N^{-1} \Delta t})[k]  + \sum_{m= n}^N  (\theta_{s_m} \star \varepsilon_{s_m^{-2} \Delta t \Delta s_m})[k \Delta t] 
$$
and therefore the Eq. \eqref{eq:def_X_d} can be implemented as
described below:

\begin{algorithm}[H]
	\label{x_algo}
	\caption{Generate a sample of $X_\ell[k \Delta t]$}
	
	\begin{algorithmic}
		\State $s \leftarrow s_N$
		\State $\Delta s \leftarrow s_N-s_{N-1}$
		\State {\bf Generate} $\varepsilon_{s^{-1} \Delta t}[k]$
		\State $\omega[k] \leftarrow (\theta_{s} \star \varepsilon_{s^{-1} \lambda^2 \Delta t})[k]$ 
		\State $X[k] \leftarrow \Delta s \; s^{H-2} \; (\psi_s \star e^{\omega})[k]$
		\State $n \leftarrow N-1$
		\While{[$n > 1$]}
		 \State $s \leftarrow s_n$
		 \State $\Delta s \leftarrow s_n-s_{n-1}$
		 \State {\bf Generate} $\varepsilon_{s^{-2} \Delta t \Delta s}[k]$
		 \State $\omega[k] \leftarrow \omega[k] + (\theta_{s} \star \varepsilon_{s^{-2} \Delta t \Delta s})[k]$ 
	 	 \State $X[k] \leftarrow X[k]+\Delta s \; s^{H-2} \; (\psi_s \star e^{\omega})[k]$
	 	 \State $n \leftarrow n-1$
		\EndWhile
		\Return $X$
	\end{algorithmic}
\end{algorithm}

\section{The dissipative anomaly}
\label{App:diss_anomaly}
The dissipative anomaly is a property one expects in fully developed turbulence to 
conciliate the fact that, when the Reynolds number becomes arbitrary large (i.e.
the kinematic viscosity $\nu$ goes to zero), on one hand the gradient of 
the velocity field diverges (i.e., $v$ becomes
non-differentiable since it corresponds to a global regularity close to $H=1/3)$
while the dissipation rate $\varepsilon \sim \nu (\frac{\partial v}{\partial x})^2$ 
remains finite.
If one denotes by $\eta$ the kolmogorov scale, i.e. the scale above which $v$ is smooth,
from the behavior of the velocity increments, $(\delta_\ell v)^3 \sim \varepsilon \ell$,
one can approximate the gradient as $\frac{\partial v}{\partial x} \sim \varepsilon^{1/3} \eta^{-2/3}$
so that the previous finite dissipation rate condition holds provided
\begin{equation}
\label{eq:eta_nu}
\eta \sim \left(\frac{\nu^3}{\varepsilon}\right)^{1/4} \Leftrightarrow \nu \sim \varepsilon^{1/3} \eta^{4/3}
\end{equation}

Let us see in what respect such a dissipative anomaly can hold within our model.
For that purpose one has to seek for a "viscosity" $\nu(\ell)$ such that, when $\ell \to 0$,
$\nu(\ell) \to 0$  and there exists 
$0 < \varepsilon < \infty$ verifying
\begin{equation}
\label{eq:diss_an1}
\lim_{\ell \to 0} \nu(\ell) \; \EE{ \left(\frac{\partial X_\ell(t)}{\partial t }\right)^2 } = \varepsilon \; .
\end{equation}

Let us denote $\varphi'(t)$, the derivative of $\varphi$ by $\vartheta(t)$.  
We thus have
\begin{equation*}
		\left(\frac{\partial X_\ell(t)}{\partial t }\right)^2  = \int_{\ell}^T \int_{\ell}^T (s_1 s_2)^{H-3} ds_1 ds_2 
		\int \int  e^{\omega_{s_1}(b_1)+\omega_{s_2}(b_2)} \vartheta \left(\frac{b_1-t}{s_1} \right) \vartheta \left(\frac{b_2-t}{s_2} \right) db_1 db_2
\end{equation*}

Since $\vartheta$ is bounded and for $s_1>s_2$ and
$$
\EE{e^{\omega_{s_1}(b_1)+\omega_{s_2}(b_2)}} \leq s_1^{-\phi(2)}
$$
it results that
$$
\nu(\ell) \; \EE{ \left(\frac{\partial X_\ell(t)}{\partial t }\right)^2 } 
\leq K \nu(\ell) \int_{\ell}^T ds_2 s_2^{H-2} \int_{s_2}^T s_1^{H-2-\phi(2)} 
$$
which means that
$$
\nu(\ell) \; \EE{ \left(\frac{\partial X_\ell(t)}{\partial t }\right)^2 }
= \OO{ \nu(\ell) \ell^{2H-2-\phi(2)}}  \; .
$$
In order to reproduce the dissipative anomaly it thus suffices to choose
\begin{equation}
\label{eq:diss_ano}
\nu(\ell) \sim \ell^{2(1-H)+\phi(2)} \; .
\end{equation}

Let us see which kind of scaling Eq. \eqref{eq:diss_ano} leads to if one wants 
to fit turbulence within our framework.
One can choose a normal law of $\omega_\ell(t)$ with intermittency 
coefficient $\lambda^2 = 0.025$ \cite{ArnMuzRou97,ArnManMuzy98,chan00}, i.e. $\phi(q) = q(q-1) \lambda^2/2$ satisfying 
$\phi(1) = 0$.
The linear behavior of third order structure function leads 
to the choice $\zeta_3 = 1 = 3H-\phi(3) = 3H -3 \lambda^2$ and therefore
$H = 1/3 + \lambda^2$. It thus 
results that Eq. \eqref{eq:diss_ano} can be rewritten 
as $\nu(\ell) \sim \ell^{2-2/3 -2 \lambda^2 + \lambda^2}  \sim \ell^{4/3-\lambda^2}$
which corresponds, up to the intermittency correction $-\lambda^2$, 
to the relationship \eqref{eq:eta_nu} obtained within the Kolmogorov approach to turbulence.

\vspace*{1cm}

\section{Computation of the leverage function}
\label{App:Leverage}
The explicit expression of the leverage function \eqref{eq:def_lev} in the case of continuous wavelet cascade as defined
in Eq. \eqref{eq:def_X}  is quite intricate:
$$
{\cal L}_{q}(\tau) = Z_{q,\ell}^{-1}  \int_0^T s^{H-2} ds \int  \left[\varphi \left(\frac{-b}{s} \right)
	-\varphi \left(\frac{-b-\ell}{s} \right) \right]  
	\EE{ e^{\omega_{s}(b)} \left| \delta_\ell X(\tau) \right|^q } db
$$
where $\delta_{\ell} X(t)$ stands for $X(t)-X(t-\ell)$.
In order to study the behavior of such an expression, 
we will first make the following approximation: 
$$\EE{ e^{\omega_{s}(t_1)} | \delta_\ell X(t_2) |^q} \simeq \ell^H \EE{e^{\omega_s(t_1)+q \omega_{\ell}(t_2) }}
$$ 
This approximation is hard to establish on a rigorous ground but can be 
intuitively motivated by the fact that when factorizing $\omega_\ell(t)$
in all $\omega_s(b)$ involved in $\delta_{\ell} X(t)$, the modulus of remaining 
integral has almost vanishing correlations with $\omega_{\ell}(t')$ and $\delta_{\ell}X(t')$ for $|t'-t| > \ell$. 
We have checked numerically that when we effectively replace $|\delta_\ell X(t) |^q$ by $\ell^{qH} e^{q \omega_\ell(t)}$,
the estimations of the leverage functions are basically unchanged.
We are thus left to estimate the following integral (the factor $\ell^{qH}$ has been absorbed in the redefinition of $Z_{q,\ell}$):
\begin{equation}
\label{eq:Lev1}
	{\cal L}_{q}(\tau) = Z_{q,\ell}^{-1} \int_0^T \! \! s^{H-2} ds \int  \! db \EE{e^{\omega_{s}(b)+q\omega_{\ell}(\tau)}} \delta_{\ell \over s} \varphi \left(\frac{-b}{s} \right)
\end{equation}
Hereafter we will exclusively elaborate on the case $q=1$ but the case of arbitrary $q$ can be considered along the same lines.
Let $C_{s,\ell}(z) =  \EE{e^{\omega_{s}(b)+\omega_{\ell}(b+z)}}$. Thanks to Eqs. \eqref{charf} and \eqref{rhoexact} and given
the condition $\phi(1)=0$, we have $C_{s,\ell}(z) = C_{\max(\ell,s)}(z)$ with:
\begin{equation}
\label{Cs}
C_{s'}(z)=
\left\{
\begin{array}{ll} 
T^{\gamma} e^{\gamma}s'^{-\gamma}e^{-\gamma\frac{|z|}{s'}} & \mbox{if}~|z| \leq s' \; , \\ 
T^{\gamma} |z|^{-\gamma} & \mbox{if}~T \ge  |z| \ge s' \; , \\
1 & \mbox{if}~ |z| > T \; .
\end{array}
\right. 
\end{equation} 
where we have set $\gamma = \phi(2)$.
For the sake of simplicity and without loss of any generality
we set $T=1$ in the following. We will also consider a simpler version
of $C_s(z)$ (notably used e.g. in \cite{MuDeBa00,BaDeMu01}) that is easier to handle in 
numerical and analytical computations:
\begin{equation}
\label{Cs2}
C_{s'}(z) = (s'+|z|)^{-\gamma}
\end{equation}

Equation \eqref{eq:Lev1} can then be rewritten as:
\begin{equation*}
\label{eq:Lev2}
{\cal L}_{1}(\tau)  =  Z_{1,\ell}^{-1} \int_0^1 \! \! s^{H-1} ds \int  \! du  \varphi \left(u \right)
  \left(C_{\max(s,\ell)}(\tau+su)-C_{\max(s,\ell)}(\tau+\ell+su) \right) 
\end{equation*}
Let us decompose ${\cal L}_1(\tau)$ as ${\cal L}_1(\tau) = {\cal L}_{-}(\tau)+{\cal L}_{+}(\tau)$ where
\begin{eqnarray*}
	{\cal L}_{-}(\tau) & = & Z_{1,\ell}^{-1} \int_0^{\ell} \! \! s^{H-1} ds \int  \! du  \varphi \left(u \right) 
	 \left(C_{\ell}(\tau+su)-C_{\ell}(\tau+\ell+su) \right) \\
	{\cal L}_{+}(\tau) & = & Z_{1,\ell}^{-1} \int_{\ell}^1 \! \! s^{H-1} ds \int  \! du  \varphi \left(u \right) 
	 \left(C_{s}(\tau+su)-C_{s}(\tau+\ell+su) \right) 
\end{eqnarray*} 
Let us choose $Z_{1\ell} = Z_1 \ell$. 
In the range $s<\ell$, we can write
(because $\varphi$ is supported in $[-1,1]$)
$$
C_{\ell}(\tau+su)-C_{\ell}(\tau+\ell+su)
\simeq -su C'_\ell(\tau+\ell)
$$
Since, for $\tau > \ell$, $|C'_\ell(z)| = \OO{|\tau|^{-1-\gamma}}$,
the contribution of $s \in [0,\ell]$ in ${\cal L}_1(\tau)$ 
is therefore of order 
$$ {\cal L}_-(\tau) \sim Z_1^{-1} \ell^{-1} \tau^{-\gamma-1} \int_0^\ell s^{H} \sim Z_1^{-1} \ell^{H} \tau^{-1-\gamma} .$$
When $\ell < s$, we write:
\begin{eqnarray*}
	&& {\cal L}_{+}(\tau)  \simeq  - Z_{1}^{-1} \int_{\ell}^1 \! \! s^{H-1} ds \int  \! du  \varphi \left(u \right) C_s'(\tau+su) \\
	  & & =  - |\tau|^{H-1-\gamma} Z_{1}^{-1} \int_{\ell|\tau|^-1}^{|\tau|^{-1}} \! \! s^{H-1} ds \int  \! du  \varphi \left(u \right) C_s'(su \pm 1)
\end{eqnarray*}
where, the last expression results from the change of variable 
$s \rightarrow |\tau|^{-1} s$ and $\pm$ corresponds to the sign
of the lag $\tau$.
Let us define
(if both integrals converge):
\begin{eqnarray*}
	 C_{+}  & = &  - Z_1^{-1} \int_{0}^{+\infty} \! \! s^{H-1} ds \int  \! du  \varphi \left(u \right) C_s'(su+1) \\
	 C_{-}  & = &  - Z_1^{-1} \int_{0}^{+\infty} \! \! s^{H-1} ds \int  \! du  \varphi \left(u \right) C_s'(su-1) \\
\end{eqnarray*}
In the domain $\tau \gg \ell$, we thus have shown that:
\begin{equation}
   {\cal L}_1(\tau) \simeq C_{\pm} |\tau|^{H-1-\gamma} 
\end{equation}
where the constants $C_+$ and $C_-$ correspond to the 
ranges $\tau >0$ and $\tau<0$ respectively.
The amplitude of the "leverage effect" can thus be measured 
by the ratio:
\begin{equation}
\label{eq:def_levratio}
  \kappa = \frac{C_+}{C_-}
\end{equation}
which may strongly depend on the chosen wavelet $\varphi$.

%
%
%
%

\end{document}